\documentclass[reprint,twocolumn,pre,showpacs,amsmath,amssymb,aps]{revtex4-1}
\usepackage{natbib}	
\usepackage{graphicx,color,rotating}
\usepackage[latin1]{inputenc}
\usepackage{textcomp}
\usepackage{dcolumn}% Align table columns on decimal point
\usepackage{bm}     % bold math
\usepackage{upgreek}
\usepackage{subdepth}  			% Added by JTK (aligns sub-indices)
\usepackage[latin1]{inputenc}

\usepackage{array}
\newcolumntype{L}[1]{>{\raggedright\let\newline\\\arraybackslash\hspace{0pt}}m{#1}}
\newcolumntype{C}[1]{>{\centering\let\newline\\\arraybackslash\hspace{0pt}}m{#1}}
\newcolumntype{R}[1]{>{\raggedleft\let\newline\\\arraybackslash\hspace{0pt}}m{#1}}
\renewcommand{\vec}[1]{\bm{#1}}
\newcommand{\ee}{\mathrm{e}}
\newcommand{\ii}{\mathrm{i}}
\newcommand{\dm}{\mathrm{d}}
\newcommand{\dd}{\mathrm{d}}
\newcommand{\avr}[1]{\big\langle #1 \big\rangle}

\DeclareMathOperator{\re}{Re}
\DeclareMathOperator{\im}{Im}

\newcommand{\taup}{\tau_\mathrm{p}}

\newcommand{\ve}{\varepsilon}

\newcommand{\veac}{\ve_\mathrm{ac}}

\newcommand{\pp}{\partial}
\newcommand{\ppsqr}{\partial^{\,2_{}}}

\newcommand{\nablabf}{\boldsymbol{\nabla}}

\newcommand{\Lapl}{\nabla^2}

\newcommand{\rot}{\nablabf\times}
\newcommand{\divop}{\nablabf\cdot}
\newcommand{\ppt}{\partial_t}

\newcommand{\dpst}{\displaystyle}

\newcommand{\fracsmall}[2]{\mbox{$\frac{#1}{#2}$}}

\newcommand{\FFFrad}{\vec{F}^\mathrm{rad}}

\newcommand{\kc}{k_\mathrm{c}}
\newcommand{\kt}{k_\mathrm{t}}
\newcommand{\ks}{k_\mathrm{s}}

\newcommand{\nnn}{\vec{n}}

\newcommand{\PPP}{\vec{P}}

\newcommand{\rrr}{\vec{r}}

\newcommand{\uuuT}{\vec{u}^{{}}_\mathrm{T}}
\newcommand{\uuuL}{\vec{u}^{{}}_\mathrm{L}}
\newcommand{\uuu}{\vec{u}}

\newcommand{\vvv}{\vec{v}}

\newcommand{\vp}{v_\mathrm{p}}

\newcommand{\zerovec}{\boldsymbol{0}}

\newcommand{\cpTi}{\tilde{c}_p}

\newcommand{\cp}{c_p}
\newcommand{\CV}{C_V}
\newcommand{\cV}{c_V}

\newcommand{\Dth}{D_\mathrm{th}}
\newcommand{\DthTi}{\tilde{D}_\mathrm{th}}

\newcommand{\Eac}{E_\mathrm{ac}}

\newcommand{\kth}{k_\mathrm{th}}
\newcommand{\kthTi}{\tilde{k}_\mathrm{th}}

\newcommand{\kapT}{\kappa_T}
\newcommand{\kapS}{\kappa_s}

\newcommand{\kapsTi}{\tilde{\kappa}_s}

\newcommand{\alfpTi}{\tilde{\alpha}_p}

\newcommand{\delt}{\delta_\mathrm{t}}

\newcommand{\dels}{\delta_\mathrm{s}}

\newcommand{\etaB}{\eta^\mathrm{b}}

\newcommand{\etaO}{\eta_0}

\newcommand{\etaOTi}{\tilde{\eta}_0}

\newcommand{\Gamt}{\Gamma_\mathrm{t}}
\newcommand{\Gams}{\Gamma_\mathrm{s}}

\newcommand{\fO}{f_0}
\newcommand{\fOfl}{f_0^\mathrm{fl}}
\newcommand{\fOsl}{f_0^\mathrm{sl}}
\newcommand{\fI}{f_1}
\newcommand{\fIfl}{f_1^\mathrm{fl}}
\newcommand{\fIsl}{f_1^\mathrm{sl}}

\newcommand{\gO}{g_0}
\newcommand{\gI}{g_1}

\newcommand{\kO}{k_0}

\newcommand{\pI}{p_1}

\newcommand{\sI}{s_1}

\newcommand{\TO}{T_0}
\newcommand{\TI}{T_1}

\newcommand{\vvvO}{\vvv_0}

\newcommand{\vvvI}{\vvv_1}

\newcommand{\vvvII}{\vvv_2}

\newcommand{\rhoO}{\rho_0}
\newcommand{\rhoI}{\rho_1}
\newcommand{\rhoII}{\rho_2}

\newcommand{\rhoTi}{\tilde{\rho}}
\newcommand{\rhoOTi}{\tilde{\rho}_0}

\newcommand{\SIkHz}{\textrm{kHz}}

\newcommand{\SImum}{\textrm{\textmu{}m}}

\newcommand{\SInm}{\textrm{nm}}

\newcommand{\beq}[1]{\begin{equation} \eqlab{#1}}
\newcommand{\eeq}{\end{equation}}
\newcommand{\bsub}{\begin{subequations}}
\newcommand{\esub}{\end{subequations}}
\def\bal#1\eal{\begin{align}#1\end{align}}
\def\bsubal#1\esubal{\bsub \begin{align}#1\end{align} \esub}
\newcommand{\nn}{\nonumber}
\newcommand{\eqlab}[1]{\label{eq:#1}}
\renewcommand{\eqref}[1]{Eq.~(\ref{eq:#1})}
\newcommand{\eqrefnoEq}[1]{(\ref{eq:#1})}
\newcommand{\eqsref}[2]{Eqs.~(\ref{eq:#1}) and~(\ref{eq:#2})}
\newcommand{\eqssref}[3]{Eqs.~(\ref{eq:#1}), (\ref{eq:#2}) and~(\ref{eq:#3})}

\newcommand{\figref}[1]{Fig.~\ref{fig:#1}}

\newcommand{\figlab}[1]{\label{fig:#1}}
\newcommand{\appref}[1]{Appendix~\ref{sec:#1}}

\newcommand{\secref}[1]{Section~\ref{sec:#1}}

\newcommand{\seclab}[1]{\label{sec:#1}}
\newcommand{\tabref}[1]{Table~\ref{tab:#1}}

\newcommand{\tablab}[1]{\label{tab:#1}}

\newcommand{\Gamv}{\Gamma_{\mathrm{s}}}

\newcommand{\vpsi}{\vec{\psi}}

\newcommand{\psis}{\psi_\mathrm{s}}
\newcommand{\phic}{\phi_\mathrm{c}}
\newcommand{\phit}{\phi_\mathrm{t}}

\newcommand{\bt}{b_\mathrm{t}}
\newcommand{\bc}{b_\mathrm{c}}

\newcommand{\btp}{b_\mathrm{t}^{\prime}}
\newcommand{\bcp}{b_\mathrm{c}^{\prime}}

\newcommand{\xc}{x_\mathrm{c}}
\newcommand{\xs}{x_\mathrm{s}}
\newcommand{\xt}{x_\mathrm{t}}
\newcommand{\xcp}{x_\mathrm{c}^{\prime}}
\newcommand{\xsp}{x_\mathrm{s}^{\prime}}
\newcommand{\xtp}{x_\mathrm{t}^{\prime}}
\newcommand{\xcpsqr}{x_\mathrm{c}^{\prime 2}}
\newcommand{\xspsqr}{x_\mathrm{s}^{\prime 2}}

\newcommand{\Phiac}{\Phi_\mathrm{ac}}
\newcommand{\dij}{\delta_{ij}}
\newcommand{\sigmabf}{\bm{\sigma}}
\newcommand{\cL}{c_\mathrm{L}}
\newcommand{\cT}{c_\mathrm{T}}
\newcommand{\cTp}{c'_\mathrm{T}}
\newcommand{\cTpsqr}{c'^{2}_\mathrm{T}}

\newcommand{\uuuI}{\vec{u}_1}

\newcommand{\chip}{\chi^\prime}
\newcommand{\chiTi}{\tilde{\chi}}

\newcommand{\sigmabfI}{\bm{\sigma}^{{}}_1}
\newcommand{\sigmabfII}{\bm{\sigma}^{{}}_2}

\begin{document}
%\preprint{Preprint identifier}

\title{Forces acting on a small particle in an acoustical field in a thermoviscous fluid}

\author{Jonas Tobias Karlsen}
\email{jonkar@fysik.dtu.dk}
\affiliation{Department of Physics, Technical University of Denmark, DTU Physics Building 309, DK-2800 Kongens Lyngby, Denmark}

\author{Henrik Bruus}
\email{bruus@fysik.dtu.dk}
\affiliation{Department of Physics, Technical University of Denmark, DTU Physics Building 309, DK-2800 Kongens Lyngby, Denmark}

\date{3 July 2015, submitted to Phys.~Rev.~E}

\begin{abstract}
We present a theoretical analysis of the acoustic radiation force on a single small particle, either a thermoviscous fluid droplet or a thermoelastic solid particle, suspended in a viscous and heat-conducting fluid medium. Our analysis places no restrictions on the length scales of the viscous and thermal boundary layer thicknesses $\dels$ and $\delt$ relative to the particle radius $a$, but it assumes the particle to be small in comparison to the acoustic wavelength $\lambda$. This is the limit relevant to scattering of sound and ultrasound waves from micrometer-sized particles. For particles of size comparable to or smaller than the boundary layers, the thermoviscous theory leads to profound consequences for the acoustic radiation force. Not only do we predict forces orders of magnitude larger than expected from ideal-fluid theory, but for certain relevant choices of materials, we also find a sign change in the acoustic radiation force on different-sized but otherwise identical particles. This phenomenon may possibly be exploited in handling of submicrometer-sized particles such as bacteria and vira in lab-on-a-chip systems.
\end{abstract}

% \pacs{43.25.Qp, 43.20.Fn, 43.20.+g, 47.35.Rs}

% 43 Acoustics
%   43.20.+g: General linear acoustics
%   43.20.Fn: Scattering of acoustic waves
%   43.20.Ks: Standing waves, resonance, normal modes
%	43.25.Nm: Acoustic streaming
%   43.25.Qp: Radiation pressure
%	43.80.-n, 43.80.+p: Ultrasound application to biology
% 47 Fluid dynamics
%   47.15.-x: laminar
%	47.35.Rs: Sound waves in fluids

%\keywords{Suggested keywords} Use showkeys class to display keywords

\maketitle

% Main text

\section{Introduction}
\seclab{Intro}

The acoustic radiation force is the time-averaged force exerted on a particle in an acoustical field due to scattering of the acoustic waves from the particle. Theoretical studies of the acoustic radiation force date back to King in 1934 \cite{King1934} and Yosioka and Kawasima in 1955 \cite{Yosioka1955}, who considered the force on an incompressible and a compressible particle, respectively, in an inviscid ideal fluid. Their work was summarized and generalized in 1962 by Gorkov \cite{Gorkov1962}, however, with the analysis still limited to ideal fluids and valid only for particles with a radius $a$ much smaller than the acoustic wavelength $\lambda$.

In subsequent work, Doinikov developed general theoretical schemes for calculating acoustic radiation forces including viscous and thermoviscous effects \cite{Doinikov1997a, Doinikov1997b, Doinikov1997c}. The direct applicability of these studies is hampered by the generality of the developed formalism, and analytical expressions are given only in the special limits of $\delta \ll a \ll \lambda$ and $a \ll \delta \ll \lambda$, where $\delta$ is the boundary layer thicknesses. Similarly, the work of Danilov and Mironov, including viscous effects, only provides analytical expressions in these two limits~\cite{Danilov2000}. However, micrometer-sized particles at kHz or MHz frequency relevant to acoustic levitation \cite{Brandt2001, Xie2001, Vandaele2005, Foresti2014} and lab-on-a-chip applications \cite{Bruus2011, Barnkob2010, Thevoz2010, Augustsson2011, Grenvall2009, Liu2012, Hammarstrom2012, Scmid2014, Antfolk2014, Carugo2014, Shields2014, Leibacher2015, Peng2015} are outside these limits, because then $\delta \sim a \ll \lambda$. This more general case was subsequently studied analytically by Settnes and Bruus including viscous boundary layers of arbitrary size~\cite{Settnes2012}.

In this work we extend the radiation force theory for droplets and elastic particles to include the effect of both viscosity and heat conduction, thus accounting for the viscous and thermal boundary layers of thickness $\dels$ and $\delt$, respectively, and we give closed-form analytical expressions in the limit of $\dels , \delt , a \ll \lambda$ with no further restrictions between $\dels$, $\delt$, and $a$. Our approach to the full thermoviscous scattering problem follows that of Epstein and Carhart from 1953 \cite{Epstein1953}. The scope of their work was a theory for the absorption of sound in emulsions such as water fog in air. In 1972, Allegra and Hawley further developed the theory to include elastic solid particles suspended in a fluid in order to calculate attenuation of sound in suspensions and emulsions \cite{Allegra1972}. The seminal work of these authors have become known as ECAH theory within the field of ultrasound characterization of emulsions and suspensions, and combined with the multiple wave scattering theories of Refs.~\cite{Foldy1945, Lloyd1967} it has been applied to calculate homogenized complex wavenumbers of suspensions and emulsions \cite{McClements1989, Challis2005}.

\begin{figure*}[!!t]
\centering
\includegraphics[width=0.75\textwidth]{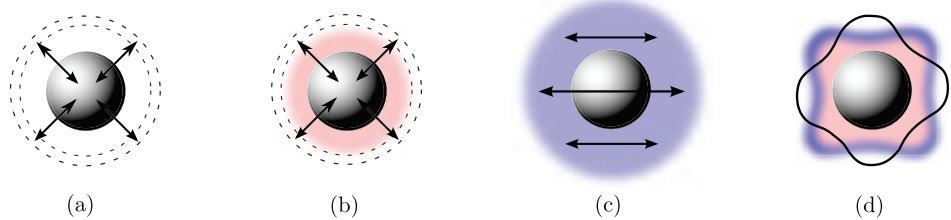}
\caption[]{(Color online)
Sketches of the physical mechanisms responsible for various multipole components in the scattering of an incident acoustic wave on a particle. (a) Compressibility contrast: the incident periodic pressure field compresses the particle relative to the fluid, which leads to monopole radiation. (b) Thermal contrast: the incident periodic temperature field leads to thermal expansion of the particle relative to the fluid, which also gives rise to monopole radiation and the development of a diffusive thermal boundary layer (pink). (c) Density contrast: a difference in inertia between particle and fluid causes the particle to oscillate relative to the fluid, which gives rise to dipole radiation and the development of a viscous boundary layer (blue). (d) Particle resonances: acoustic wavelengths comparable to the particle size leads to complex shape changes, which gives rise to  multipole radiation and a complex thermoviscous boundary layer (pink/blue).}
\figlab{rad_force_concept}
\end{figure*}

The field of ultrasound characterization driven by engineering applications and the field of acoustic radiation forces have developed in parallel with little overlap. Indeed, the scopes of the work in the two fields are very different. In the works of Epstein and Carhart and Allegra and Hawley, there is no mention of acoustic radiation forces \cite{Epstein1953,Allegra1972}. However, the underlying scattering problem of a particle suspended in a fluid remains the same, and having once solved for the amplitude of the propagating scattered wave, the acoustic radiation force on the particle may be obtained from a far-field calculation. In the far field, the propagating scattered field changes, when taking into account the thermoviscous scattering mechanisms, including boundary layer losses and excitation of acoustic streaming in the vicinity of the particle. In this work we will elucidate this approach, as it leads to a particularly simple and valuable formulation for the acoustic radiation force in the long-wavelength limit \cite{Settnes2012}.

Considering the success of the ECAH method to describe attenuation of sound in emulsions and suspensions, we can with great confidence apply the method to analyze the consequences of thermoviscous scattering on the acoustic radiation force. Nevertheless, we find a need to re-examine the problem of thermoviscous scattering in order to apply the theory to the problem of acoustic radiation forces in a clear and consistent manner. One point of clarification relates to an ambiguity in the thermoelastic solid theory presented by Allegra and Hawley~\cite{Allegra1972}, where no clear distinction is made between isothermal and adiabatic solid parameters, thus tacitly implying $\gamma=\cp/\cV=1$ in solids. Here, we will provide a self-consistent treatment of thermoviscous scattering that clarifies this issue and allows for ease of comparison with existing acoustic radiation force theories.

\begin{table}[!b]
\caption{\tablab{eqs_overview} References to analytical expressions derived in this paper for the monopole and dipole scattering coefficients $\fO$ and $\fI$ in the long-wavelength limit $a \ll \lambda$. For any given incident acoustic field, the acoustic radiation force $\FFFrad$ is calculated using \eqref{Frad_Settnes} with these expressions for $\fO$ and $\fI$.}
\begin{ruledtabular}
\begin{tabular}{lcc}
Size of particle and boundary layers & $f_0$ & $f_1$ \\ \hline
\multicolumn{3}{l}{\textit{Thermoviscous droplet:}\rule{0mm}{3.0ex}} \\
Arbitrary particle size & \eqref{f0_fluid-fluid}  & \eqref{f1_fluid-fluid}\\
Small-width boundary layers & \eqref{f0_fluid-fluid_weak}  & \eqref{f1_fluid-fluid_weak}\\
Zero-width boundary layers & \eqref{f0_fluid-fluid_large} & \eqref{f1_large}\\
Point-particle limit & \eqref{f0_fluid-fluid_point} & \eqref{f1_point}\\
\multicolumn{3}{l}{\textit{Thermoelastic particle:}\rule{0mm}{3.0ex}}\\
Arbitrary particle size & \eqref{f0_fluid-solid} & \eqref{f1_fluid-solid}\\
Small-width boundary layers & \eqref{f0_fluid-solid_weak} & \eqref{f1_fluid-solid_weak}\\
Zero-width boundary layers & \eqref{f0_fluid-solid_large} & \eqref{f1_large} \\
Point-particle limit & \eqref{f0_fluid-solid_point} & \eqref{f1_point}
\end{tabular}
\end{ruledtabular}
\end{table}

Before proceeding with the mathematical treatment, we refer the reader to \figref{rad_force_concept}, which illustrates the physical mechanisms responsible for the monopole, dipole, and multipole scattering from a particle subject to a periodic acoustic field \cite{Challis2005}. The final results for the acoustic radiation force are presented in terms of corrected expressions for the monopole and dipole scattering coefficients $f_0$ and $f_1$. This approach allows for an easy comparison to the ideal fluid theory and moreover, as shown by Settnes and Bruus \cite{Settnes2012}, it provides a simple way of evaluating acoustic radiation forces for any given incident acoustic field. To this end, \tabref{eqs_overview} provides an overview of the equations needed to evaluate the thermoviscous acoustic radiation force on small droplets or solid particles.

\section{Basic considerations on the acoustic radiation force}
\seclab{Frad1}

We consider a single particle or droplet suspended in an infinite, quiescent fluid medium with no net body force, but perturbed by a time-harmonic acoustic field with angular frequency $\omega$. The density, velocity, and stress of the perturbed fluid is denoted  $\rho$, $\vvv$, and $\sigmabf$, respectively. The region $\Omega(t)$ occupied by the particle, its surface $\pp\Omega(t)$, and the outward-pointing surface vector $\nnn$ depend on time due to the acoustic field. The instantaneous acoustic radiation force is given by the surface integral of the fluid stress $\sigmabf$  acting on the particle surface. However, since  the short time scale corresponding to the oscillation period $\tau$ is not resolved experimentally, we define the acoustic radiation force $\FFFrad$ in the conventional time-averaged sense \cite{King1934, Yosioka1955, Gorkov1962, Doinikov1997a, Danilov2000, Settnes2012},
 \beq{FradDef1}
 \FFFrad = \left\langle \oint_{\pp\Omega(t)} \sigmabf\cdot\nnn\:\dm a \right\rangle,
 \eeq
where the angled bracket denotes the time average over one oscillation period. Notice that this definition includes the acoustic streaming generated locally near the particle, since the stresses leading to this streaming are contained in the fluid stress tensor $\sigmabf$. In contrast, by considering an infinite domain, we are excluding effects of what Danilov and Mironov refer to as external streaming \cite{Danilov2000}, which would be generated at the boundaries of any finite domain. For a given finite domain, the external streaming can be calculated \cite{Muller2012}, and the total force acting on a particle is the sum of the radiation force and the external-streaming-induced Stokes drag. This approach has been used in studies of particle trajectories and has been validated experimentally \cite{Barnkob2012a, Muller2013}.

We consider a state, which is periodic in the acoustic oscillation period $\tau$, tantamount to requiring that any non-periodic phenomenon, such as particle drift, is negligible within one oscillation period. Usually, this requirement is not very restrictive, as discussed in more detail in \secref{constraints}. For a time-periodic state, any field can be written as a Fourier series  $f(\rrr,t) = \sum_{n=0}^\infty f_n(\rrr)\:\ee^{-\ii n \omega t}$, with $\omega = 2\pi/\tau$, and the time-average of any total time derivative is zero, $\avr{\frac{\dm}{\dm t}f(\rrr,t)} = 0$.

A useful expression for $\FFFrad$ is obtained by considering the momentum flux density $\sigmabf -\rho\vvv\vvv$ entering the fluid volume between the particle surface $\pp\Omega(t)$ and an arbitrary static surface $\pp\Omega_1$ enclosing the particle. The total momentum $\PPP$ of the fluid in this volume is the volume integral of $\rho\vvv$, and because the net body force on the fluid is  zero, the time-averaged rate of change $\avr{\frac{\dm}{\dm t}\PPP}$ is
 \bal
 \nn
 \bigg\langle\frac{\dm \PPP}{\dm t}\bigg\rangle &=
 \bigg\langle \oint_{\pp\Omega_1}\!\!\Big[
 \sigmabf - \rho\vvv\vvv\Big]\!\cdot\nnn\:\dm a\!  \bigg\rangle
 \!+\!
 \bigg\langle\! \oint_{\pp\Omega(t)}\!\!\!\!\!\!
 \sigmabf \cdot(-\nnn)\:\dm a \bigg\rangle \\
 &
 \eqlab{PrateFlux1}
 =\bigg\langle \oint_{\pp\Omega_1}\!\!\Big[
 \sigmabf - \rho\vvv\vvv\Big]\!\cdot\nnn\:\dm a\!  \bigg\rangle
 - \FFFrad.
 \eal
Here, $\nnn$ is the surface vector pointing out of $\pp\Omega_1$ (out of the fluid) and out of $\pp\Omega(t)$ (into the fluid). The advection term $\rho\vvv\vvv$ is zero at $\pp\Omega(t)$, since there is no advection of momentum through the interface of the particle. Finally, using that the time average of the total time derivative $\avr{\frac{\dm \PPP}{\dm t}}$ is zero in the time-periodic system, we obtain
\bal
\eqlab{FradExactFF}
 \FFFrad =  \bigg\langle \oint_{\pp\Omega_1}\Big[
 \sigmabf - \rho\vvv\vvv\Big]\cdot\nnn\:\dm a  \bigg\rangle.
\eal
Thus, even before applying perturbation theory, the acoustic radiation force can be evaluated as the total momentum flux through any static surface $\pp\Omega_1$ enclosing the particle. To second order in the acoustic perturbation, using the expansions $\rho = \rhoO +\rhoI +\rhoII$, $\vvv = \zerovec +\vvvI +\vvvII$, and $\sigmabf = \sigmabf_0 + \sigmabfI + \sigmabfII$, the radiation force \eqrefnoEq{FradExactFF} becomes
 \beq{FradSecondFarField1}
 \FFFrad = \oint_{\pp\Omega_1} \Big[  \avr{\sigmabfII} - \rhoO\avr{\vvvI\vvvI} \Big]\cdot\nnn\:\dm a,
 \eeq
where we have used that the time-average of the time-harmonic, first-order fields is zero.

In regions sufficiently far from acoustic boundary layers, the acoustic wave is a weakly damped propagating acoustic mode, for which viscous and thermal effects are negligible. This insight was used in Ref.~\cite{Settnes2012} to analytically integrate \eqref{FradSecondFarField1} by placing $\pp\Omega_1$ in the far field. In the long-wavelength limit, where the particle radius $a$ is assumed much smaller than the wavelength $\lambda$, i.e. for $\kO a \ll 1$ with $\kO=2\pi/\lambda$, it was shown that the acoustic radiation force may be evaluated directly from the incident first-order acoustic field and the expressions for the monopole and dipole scattering coefficients $\fO$ and $\fI$ for the suspended particle, as
 \bal
 \eqlab{Frad_Settnes}
 \FFFrad \!= - \pi a^3\! \left[\dfrac{2 \kappa_s}{3}
 \re\!\left[f_0^* p_\mathrm{in}^*\! \nablabf p_\mathrm{in} \right] - \rho_0\!
 \re\!\left[f_1^* \vec{v}_\mathrm{in}^* \!\cdot\! \nablabf \vec{v}_\mathrm{in} \right] \right].
 \eal
Here, $p_\mathrm{in}$ and $\vec{v}_\mathrm{in}$ is the incident acoustic pressure and velocity fields evaluated at the particle position, the asterisk denotes complex conjugation, and $\kappa_s$ and $\rho_0$ are the isentropic compressibility and the mass density of the fluid medium, respectively.

Equation~(\ref{eq:Frad_Settnes}) is valid for any incident time-harmonic, acoustic field, and consequently the problem of calculating the radiation force on a small particle reduces to calculating the coefficients $f_0$ and $f_1$. Closed, analytical expressions for these are given in the literature for small particles in the special cases of compressible particles in ideal fluids \cite{Yosioka1955, Gorkov1962} and compressible particles in viscous fluids \cite{Settnes2012}. Moreover, $f_0$ and $f_1$ can be extracted from Ref.~\cite{Doinikov1997b, Doinikov1997c} for rigid spheres and liquid droplets in thermoviscous fluids for the limiting cases of very thin and very thick boundary layers. The main result of this paper is the derivation of analytical expressions for  $f_0$ and $f_1$ for a spherical thermoviscous droplet and a thermoelastic particle suspended in a thermoviscous fluid without restrictions on the boundary layer thicknesses, see \tabref{eqs_overview}. Moreover, we provide an analysis of how $\FFFrad$ is affected by thermoviscous effects in these cases.

Finally, we note that since $\fO$ and $\fI$ depend only on frequency and material parameters, expression~\eqrefnoEq{Frad_Settnes} for the radiation force remains valid for any incident wave composed of plane waves at the same frequency. In the case of a superposition  $p_\mathrm{in} = \sum_{j=1}^N p_j(\omega_j)$ of acoustic fields $p_j(\omega_j)$ (and similarly for $\vec{v}_\mathrm{in}$) at different frequencies $\omega_j$, the resulting radiation force is obtained by summing over the forces obtained from \eqref{Frad_Settnes} for each frequency,
 \bal
 \eqlab{Frad_Settnes_freq}
 \FFFrad = - \pi a^3 \sum_{j=1}^N \Big[
 \dfrac{2 \kappa_s}{3}
 &\re\big[ f_0^*(\omega_j)\: p_j^* \nablabf p_j\big]
 \nn \\
 - \rho_0
 &\re\big[f_1^*(\omega_j)\: \vec{v}_j^* \cdot \nablabf \vec{v}_j \big] \Big].
 \eal
This generalization of \eqref{Frad_Settnes} provides a way to evaluate the acoustic radiation force on a single particle regardless of the complexity of the incident field.

\section{Thermoviscous perturbation theory of acoustics in fluids}
\seclab{Theory}

The starting point of the theory is the first law of thermodynamics and the conservation of mass, momentum, and energy.  Introducing the thermodynamic variables temperature $T$, pressure $p$, density $\rho$, internal energy $\ve$ per mass unit, entropy $s$ per mass unit, and volume per mass unit $1/\rho$, the first law of thermodynamics with $s$ and $\rho$ as independent variables becomes
 \beq{FirstLaw}
 \dm \ve = T\:\dm s - p\: \dm \bigg(\!\frac{1}{\rho} \bigg)
 = T\: \dm s + \frac{p}{\rho^2}\: \dm\rho.
 \eeq
For acoustic wave propagation it is often convenient to use $T$ and $p$ as independent thermodynamic variables. This is obtained by a Legendre transformation of the internal energy $\ve$ per unit mass to the Gibbs free energy $g$ per unit mass, $g = \ve -Ts + p\:\frac{1}{\rho}$.

Besides the first law of thermodynamics, the governing equations of thermoviscous acoustics requires the introduction of the velocity field $\vvv$ and the stress tensor $\bm{\sigma}$ of the fluid. The latter can be expressed in terms of $\vvv$, $p$, the dynamic shear viscosity $\eta$, the bulk viscosity $\etaB$, and the viscosity ratio $\beta=\etaB / \eta + 1/3$, as
 \bsubal
 \eqlab{visc_stress_tensor}
 \bm{\sigma} &= -p\:\textbf{I} + \bm{\tau}, \\
 \bm{\tau} &= \eta\bigg[\nablabf\vvv + (\nablabf \vvv)^\mathrm{T}\bigg]
 + (\beta-1) \eta \left( \divop \vvv \right) \: \textbf{I}.
 \esubal
Here, $\textbf{I}$ is the unit tensor and the superscript "T" indicates tensor transposition. The tensor $\bm{\tau}$ is the viscous part of the stress tensor assuming a Newtonian fluid \cite{Bruus2008}.

Considering the fluxes of mass, momentum and energy into a small test volume, we use Gauss's theorem to formulate the general governing equations for conservation of mass, momentum and energy in the fluid under the assumption of no net body forces and no heat sources,
 \bsub
 \eqlab{governing_equations}
 \bal
 \eqlab{contEq}
 \pp_t \rho &= \nablabf\cdot\big[-\rho\vvv\big] , \\
 \eqlab{momentumEq}
 \pp_t (\rho\vvv) &= \nablabf\cdot\big[\bm{\sigma} - \rho\vvv\vvv\big] , \\
 \eqlab{energyEq}
 \pp_t \big(\rho\ve + \fracsmall{1}{2}\rho v^2\big) &=
 \nablabf\cdot\big[\vvv\cdot\bm{\sigma} + \kth\nablabf T - \rho(\ve+\fracsmall{1}{2}v^2)\vvv\big].
 \eal
 \esub
Here, we have introduced the thermal conductivity $\kth$ assuming the usual linear form for the heat flux given by Fourier's law of heat conduction.

\subsection{First-order equations for fluids}
The zeroth-order state of the fluid is quiescent, homogeneous, and isotropic. Then, treating the acoustic field as a perturbation of this state in the acoustic perturbation parameter $\veac$, given by
 \beq{veac_def}
 \veac = \frac{|\rhoI|}{\rhoO}  \ll 1,
 \eeq
we expand all fields as $g = \gO + \gI$, but with $\vvvO = \zerovec$. The zeroth-order terms drop out of the governing equations, while the first-order mass, momentum, and energy equations obtained from \eqsref{FirstLaw}{governing_equations} become
\bsub
\eqlab{first_order_eq}
\bal
\pp_t \rho_1 &= - \rho_0 \divop \vvvI , \eqlab{first_cont} \\
\rho_0 \pp_t \vvvI &= - \nablabf p_1  + \eta_0 \Lapl \vvvI + \beta \eta_0 \nablabf \left( \divop \vvvI \right) , \eqlab{first_NS} \\
\rho_0 T_0 \pp_t s_1 &= \kth \Lapl T_1. \eqlab{first_heat}
\eal
\esub
It will prove useful to eliminate the variables $\pI$, $\rhoI$, and $\sI$ to end up with only two equations for the variables $\vvvI$ and $\TI$. To this end, we combine \eqref{first_order_eq} with the two thermodynamic equations of state $\rho=\rho(p,\,T)$ and $s=s(p,\,T)$. The total differentials of $\rhoI$ and $\sI$ are
\bsub
\eqlab{thermodynamic_variations}
\bal
\mathrm{d}\rho &= \left(\dfrac{\pp \rho}{\pp p} \right)_T \mathrm{d}p + \left( \dfrac{\pp \rho}{\pp T} \right)_p \mathrm{d}T , \\
\mathrm{d}s &= \left(\dfrac{\pp s}{\pp p}\right)_T \mathrm{d}p + \left(\dfrac{\pp s}{\pp T}\right)_p \mathrm{d}T ,
\eal
\esub
which may be linearized so that the partial derivatives of $\rho$ and $s$ refer to the unperturbed state of the fluid. This leads to the introduction of the isothermal compressibility $\kappa_T$, the isobaric thermal expansion coefficient $\alpha_p$, and the specific heat capacity at constant pressure $\cp$,
\bal
\kappa_T = \dfrac{1}{\rho} \left( \dfrac{\pp \rho}{\pp p} \right)_T , \ \alpha_p = - \dfrac{1}{\rho} \left( \dfrac{\pp \rho}{\pp T} \right)_p , \ \cp = T \left( \dfrac{\pp s}{\pp T} \right)_p .
\eal
Moreover, $(\pp s / \pp p)_T = - \alpha_p / \rho$, which may be derived as a Maxwell relation differentiating $g$ after $p$ and $T$. Thus, the linearized form of \eqref{thermodynamic_variations} is
\bsub
\eqlab{thermodynamic_identities}
\bal
\rho_1 &= \rho_0 \: \kappa_T \: p_1 - \rho_0 \: \alpha_p \: T_1 , \\
s_1 &= \dfrac{\cp}{T_0} \: T_1 - \dfrac{\alpha_p}{\rho_0} \: p_1 . \eqlab{thermodynamic_identities_s1}
\eal
\esub
We further introduce the isentropic compressibility $\kappa_s$ and the specific heat capacity at constant volume $\cV$,
\bsub
\bal
\kappa_s = \dfrac{1}{\rho} \left( \dfrac{\pp \rho}{\pp p} \right)_s,
\qquad
\cV = T \left( \dfrac{\pp s}{\pp T} \right)_V .
\eal
\esub
Then the following two well-known thermodynamic identities may be derived \cite{Landau1980},
 \beq{gamma_small}
 \kappa_T = \gamma \kappa_s , \qquad
 \gamma \equiv \dfrac{\cp}{\cV} = 1 + \dfrac{\alpha_p^2 T_0}{\rho_0 \cp \kappa_s} .
 \eeq

To proceed with the reduction of \eqref{first_order_eq}, we first  differentiate \eqref{first_NS} with respect to time and substitute $\Lapl\vvvI = \nablabf(\divop\vvvI) - \rot\rot\vvvI$. Then \eqref{thermodynamic_identities} is used to eliminate $p_1$ and $s_1$ in \eqsref{first_NS}{first_heat}, followed by elimination of $\pp_t\rhoI$ using \eqref{first_cont}. The resulting equations for $\vvvI$ and $\TI$ are
\bsub
\eqlab{first_order_eq_2}
\bal
&\ppsqr_t \vvvI - \left( \dfrac{1}{\rho_0 \kapT} + (1+\beta) \nu_0 \pp_t \right) \nablabf (\divop \vvvI) \nn \\
&\qquad\qquad + \nu_0 \pp_t \rot\rot\vvvI = - \dfrac{\alpha_p}{\rho_0 \kappa_T} \pp_t \nablabf T_1 , \eqlab{first_order_eq_2_a} \\
&\gamma \Dth \Lapl T_1 - \pp_t T_1 = \dfrac{\gamma-1}{\alpha_p}\divop\vvvI, \eqlab{first_order_eq_2_b}
\eal
\esub
where we have introduced the momentum diffusion constant $\nu_0$ and the thermal diffusion constant $\Dth$,
 \beq{nu0DthDef}
 \nu_0 = \frac{\eta_0}{\rho_0}, \qquad
 \Dth = \frac{\kth}{\rhoO\cp}.
 \eeq

\subsection{Potential equations for fluids}
The velocity field $\vvvI$ is decomposed into the gradient of a scalar potential $\phi$ (the longitudinal component) and the rotation of a divergence-free vector potential $\vpsi$ (the transverse component),
 \beq{Helmholtz_decomp}
 \vvvI = \nablabf \phi + \rot \vpsi,\; \text{ with } \divop\vpsi = 0 .
 \eeq
Inserting this well-known Helmholtz decomposition into \eqref{first_order_eq_2_a} leads to the equation
 \bal
 &\nablabf \left[ \pp_t^2 \phi - \left( \dfrac{1}{\rho_0 \kapT} + (1+\beta) \nu_0 \pp_t \right) \Lapl \phi + \dfrac{\alpha_p}{\rho_0 \kappa_T} \pp_t T_1 \right] \nn \\
 &\qquad\qquad = \rot \left[ - \pp_t^2 \vpsi + \nu_0 \pp_t \Lapl \vpsi \right] .
 \eal
In general, both sides of the equation must vanish separately, which leads to two equations. Combining these with \eqref{first_order_eq_2_b}, into which \eqref{Helmholtz_decomp} is inserted, leads to the following form of \eqref{first_order_eq_2},
 \bsub
 \eqlab{first_order_eq_3}
 \bal
 \eqlab{first_order_eq_3_a}
 \pp_t^2 \phi &= \left(\dfrac{1}{\rho_0 \kapT} + (1+\beta) \nu_0 \pp_t \right) \Lapl \phi - \dfrac{\alpha_p}{\rho_0 \kappa_T} \pp_t T_1 ,
 \\
 \eqlab{first_order_eq_3_b}
 \pp_t T_1 &= \gamma \Dth \Lapl T_1 - \dfrac{\gamma-1}{\alpha_p} \Lapl \phi ,
 \\
 \eqlab{first_order_eq_3_c}
 \pp_t \vpsi &= \nu_0 \Lapl \vpsi .
 \eal
 \esub
In the adiabatic limit, for which $\Dth=0$, the well-known adiabatic wave equation for $\phi$ is obtained by inserting \eqref{first_order_eq_3_b} into \eqrefnoEq{first_order_eq_3_a}, from which the adiabatic speed of sound $c$ for longitudinal waves is deduced,
 \beq{c_rho0_kaps}
 c = \dfrac{1}{\sqrt{\rho_0 \kappa_s}} .
 \eeq
In the isothermal case, for which $T_1=0$, the wave equation \eqrefnoEq{first_order_eq_3_a} instead describes waves traveling at the isothermal speed of sound $c/\sqrt{\gamma}=1/\sqrt{\rho_0 \kappa_T}$. For ultrasound acoustics, sound propagation in the bulk of a fluid is generally very close to being adiabatic.

\section{Thermoelastic theory of acoustics in isotropic solids}
A thermoelastic solid may be deformed by the action of applied forces or on account of thermal expansion. Following Landau and Lifshitz \cite{Landau1986}, we describe the deformation of a solid elastic body using the displacement field $\uuu$, which describes the displacement $\uuu(\vec{r},t)$ of a solid element away from its initial, undeformed position $\vec{r}$ to its new temporary position $\vec{r} + \uuu(\vec{r},t)$. Any displacement away from equilibrium gives rise to internal stresses tending to return the body to equilibrium. These forces are described using the stress tensor $\sigmabf$, which leads to the force density $\divop\sigmabf$. In the description of the thermodynamics of solids, it is advantageous to work with per-volume quantities denoted by uppercase letters, in contrast to the per-mass quantities given by lowercase letters. The first law of thermodynamics reads
\bal
\dd\mathcal{E} = T \dd S + \sigma_{ij} \dd u_{ij},
\eal
where $\mathcal{E}$ is the internal energy per unit volume, $S$ is the entropy per unit volume, and $T$ is the temperature. The work done by the internal stresses per unit volume is equal to $-\sigma_{ij} \dd u_{ij}$, where we have introduced the strain tensor $u_{ij}$, which for small displacements is given by
\bal
u_{ij} = \dfrac{1}{2} \left[ \pp_i u_j + \pp_j u_i \right] .
\eal
Transforming the internal energy per unit volume $\mathcal{E}$ to the Helmholtz free energy per unit volume  $F=\mathcal{E} - T S$, where temperature $T$ and strain $u_{ij}$ are the independent variables, the first law becomes $\dd F = - S \dd T + \sigma_{ij} \dd u_{ij}$.

Consider the undeformed state of an isotropic, thermoelastic solid at temperature $T_0$ in the absence of external forces. The free energy $F$ is then given as an expansion in powers of the temperature difference $T-\TO$ and the strain tensor $u_{ij}$. To linear order, the stress tensor $\sigma_{ij} = \big( \frac{\pp F}{\pp u_{ij}} \big)_T$ and the entropy $S = - \big( \frac{\pp F}{\pp T} \big)_{u_{ij}}$ become
 \bsubal
 \eqlab{stress_tensor_sl}
 \sigma_{ij} &= - \frac{\alpha_p(T-T_0)}{\kapT} \dij +
 \frac{E}{1+\sigma} \left[ u_{ij} + \dfrac{\sigma}{1-2\sigma}u_{kk} \dij \right], \\
 \eqlab{solid_entropy}
 S(T) &= S_0(T) + \dfrac{\alpha_p}{\kapT} u_{kk},
 \esubal
where $S_0(T)$ is the entropy of the undeformed state at temperature $T$, while $E$ and $\sigma$ are the isothermal Young's modulus and Poisson's ratio, respectively. The isothermal compressibility $\kapT$ of the solid is given in terms of $E$ and $\sigma$ as,
\bal
\eqlab{kapT_sl}
\kapT = \dfrac{3(1-2\sigma)}{E} .
\eal

\subsection{Linear equations for solids}

In elastic solids advection of momentum and heat cannot occur, so the momentum equation in the absence of body forces takes the linear form
$\rho \pp_t^2 \uuu = \divop \sigmabf$. Assuming the material parameters $\alpha_p$, $\kapT$, $E$, and $\sigma$ to be constant, it becomes
\bal
\nn
\rho \pp_t^2 \uuu &= - \frac{\alpha_p}{\kapT}\! \nablabf T + \frac{E}{2(1+\sigma)} \left[ \Lapl \uuu
+ \frac{1}{1-2\sigma} \nablabf(\divop \uuu) \right] \\
\eqlab{solid_momentum}
& = - \frac{\alpha_p}{\rho \kapT} \nablabf T\! + \cT^2 \Lapl \uuu\! + (\cL^2 \!-\! \cT^2) \nablabf (\divop \uuu),
\eal
where we have introduced the isothermal speed of sound of longitudinal waves $\cL$ and of transverse waves~$\cT$,
 \bsub
 \beq{cL_cT}
 \cL^2 = \dfrac{(1-\sigma)}{(1+\sigma) (1-2\sigma)} \dfrac{E}{\rho} , \ \ \ \cT^2 = \dfrac{1}{2(1+\sigma)}\dfrac{E}{\rho}.
 \eeq
Using the decomposition $\uuu = \uuuT + \uuuL$ in the transverse and longitudinal displacements $\uuuT$ and $\uuuL$ with $\divop \uuuT=0$ and $\rot \uuuL = \vec{0}$, respectively, it immediately follows from \eqref{solid_momentum} that in the isothermal case, transverse and longitudinal waves travel at the speed $\cT$ and $\cL$, respectively. Combining \eqsref{kapT_sl}{cL_cT} one obtains an important relation connecting the isothermal compressibility $\kapT$ of the solid to the isothermal sound speeds $\cL$ and $\cT$,
 \beq{kapT_of_cL_cT}
 \dfrac{1}{\rho \kapT} = \cL^2 - \dfrac{4}{3}\cT^2  .
 \eeq
 \esub

Turning to the energy equation, the amount of heat absorbed per unit time per unit volume is $T (\pp_t S)$. If there are no heat sources in the bulk, the rate of heat absorbed is given by the influx $-\kth \nablabf T$ of heat by conduction, and the heat equation thus becomes
 \beq{heat_general_sld}
 T (\pp_t S) = - \divop \left[ - \kth \nablabf T \right] = \kth \Lapl T,
 \eeq
where the heat conductivity $\kth$ is taken to be constant. We rewrite this equation using expression~\eqrefnoEq{solid_entropy} for the entropy, and using that the time-derivative of $S_0$ may be written as
\bal
\dfrac{\pp S_0}{\pp t} = \left(\dfrac{\pp S_0}{\pp T}\right)_{\!V} \dfrac{\pp T}{\pp t} = \dfrac{\CV}{T} \dfrac{\pp T}{\pp t} ,
\eal
where the heat capacity $\CV$ per unit volume at constant volume enters through the relation $\CV =T \:(\pp S_0 / \pp T)_V $ with the derivative taken for the undeformed state at constant volume, that is for $u_{kk}=\divop \uuu = 0$. Combining these considerations with the identity for $\gamma$ equivalent to \eqref{gamma_small}, the heat equation~\eqrefnoEq{heat_general_sld} becomes
\bal
\CV \pp_t T + \dfrac{(\gamma - 1) \CV}{\alpha_p} \pp_t \divop \uuu = \kth \Lapl T .
\eal
Finally, having eliminated all extensive thermodynamic variables, we return to per-mass quantities, such as $\cV = \CV/\rho$, and thus arrive at the coupled equations for thermoelastic solids,
\bsub
\eqlab{solid_uuu_eqs}
\bal
&\pp_t^2 \uuuI  -  \cL^2 \nablabf(\nablabf \cdot \uuuI)
 +  \cT^2\nablabf \times \nablabf \times \uuuI  = - \dfrac{\alpha_p}{\rhoO \kappa_T} \nablabf \TI , \eqlab{solid_uuu_eqs_a} \\
&\gamma \Dth \Lapl \TI - \pp_t \TI = \dfrac{\gamma-1}{\alpha_p} \pp_t \divop \uuuI , \eqlab{solid_uuu_eqs_b}
\eal
\esub
with $\gamma$ and $\Dth$ defined in \eqsref{gamma_small}{nu0DthDef}, and the linearity emphasized by the addition of subscripts "1" to the field variables. In this form, the thermoelastic equations~\eqrefnoEq{solid_uuu_eqs} correspond to the fluid equations~\eqrefnoEq{first_order_eq_2}.

\subsection{Potential equations for solids}
The time derivative $\pp_t\uuuI$ of the displacement field $\uuuI$  describes the velocity field in the solid. Analogous to the fluid case, we make a Helmholtz decomposition of this velocity field in terms of the velocity potentials $\phi$ and $\vpsi$
 \beq{dtu1}
 \ppt \uuuI = \nablabf \phi + \rot \vpsi , \ \text{with} \ \ \divop\vpsi = 0 .
 \eeq
Inserting this into \eqref{solid_uuu_eqs} and following the procedure leading to \eqref{first_order_eq_3} for fluids, we obtain the corresponding three equations for solids,
\bsub
\eqlab{solid_potential_eqs}
\bal
\pp_t^2 \phi &= \cL^2 \Lapl \phi - \dfrac{\alpha_p}{\rho \kapT} \ppt \TI , \eqlab{solid_potential_eqs_a} \\
\pp_t \TI &= \gamma \Dth \Lapl \TI - \dfrac{\gamma-1}{\alpha_p} \Lapl \phi , \eqlab{solid_potential_eqs_b} \\
\pp_t^2 \vpsi &= \cT^2 \Lapl \vpsi . \eqlab{solid_potential_eqs_c}
\eal
\esub
The main difference between the fluid and the solid case is in \eqref{solid_potential_eqs_c} for the vector potential $\vpsi$, which now takes the form of a wave equation describing transverse waves traveling at the transverse speed of sound $\cT$ instead of the diffusion equation~\eqrefnoEq{first_order_eq_3_c}.

The usual adiabatic wave equation for the scalar potential $\phi$ is obtained in the limit of $\Dth=0$ combining \eqsref{solid_potential_eqs_a}{solid_potential_eqs_b}, and the speed $c$ of adiabatic, longitudinal wave propagation in an elastic solid becomes
\bal
c^2 = \cL^2 + \dfrac{\gamma - 1}{\rhoO \kapT} . \eqlab{c_ad_solid}
\eal
For most solids $\gamma-1\ll 1$, leading to a negligible difference between the isothermal $\cL$ and the adiabatic $c$, the latter being closest to the actual speed of sound measured in ultrasonic experiments.

\section{Unified potential theory of acoustics in fluids and solids}
\seclab{unification}
The similarity between the potential equations~\eqrefnoEq{first_order_eq_3} and~\eqrefnoEq{solid_potential_eqs}, allows us to write down a unified potential theory of acoustics in thermoviscous fluids and thermoelastic solids. The main result of this section is the derivation of three wave equations with three distinct wavenumbers corresponding to three modes of wave propagation, namely two longitudinal modes describing propagating compressional waves and damped thermal waves, respectively, and one transverse mode describing a shear wave, which is damped in a fluid but propagating in a solid.

We work with the first-order fields in the frequency domain considering a single frequency $\omega$. Using complex notation, we write any first-order field $g_1(\vec{r},t)$ as
\bal
\eqlab{time-harmonic}
g_1(\vec{r},t) = g_1(\vec{r}) \mathrm{e}^{-\ii \omega t} .
\eal
Assuming this form of time-harmonic first-order fields, \eqsref{first_order_eq_3_a}{solid_potential_eqs_a} lead to expressions for the temperature field $T_1$ in a fluid (fl) and a solid (sl), respectively, in terms of the corresponding scalar potential $\phi$
\bsub
\eqlab{T1_of_potential}
\bal
T_1^\mathrm{fl} &= \dfrac{\ii \omega \rho_0 \kapT}{\alpha_p} \left[ \phi + \dfrac{c^2}{\omega^2} \dfrac{1 - \ii \gamma \Gamv }{\gamma} \Lapl \phi \right] , \eqlab{T1_of_potential_fluid} \\
T_1^\mathrm{sl} &= \dfrac{\ii \omega \rho_0 \kapT}{\alpha_p} \left[ \phi + \dfrac{\cL^2}{\omega^2} \Lapl \phi \right] . \eqlab{T1_of_potential_solid}
\eal
\esub
Here, we have introduced the dimensionless bulk damping factor $\Gamv$ accounting for viscous dissipation in the fluid. For  convenience, we also introduce the thermal damping factor $\Gamt$ accounting for dissipation due to heat conduction both in fluids and in solids. These two bulk damping factors are given by
 \bal
 \eqlab{Gamv_Gamth}
 \Gamv &= \dfrac{(1+\beta) \nu_0 \omega}{c^2} ,
 \qquad
 \Gamt = \dfrac{\Dth \omega}{c^2} .
 \eal

Substituting expression~\eqrefnoEq{T1_of_potential_fluid} for $T^\mathrm{fl}_1$ into \eqref{first_order_eq_3_b}, or expression~\eqrefnoEq{T1_of_potential_solid} for $T^\mathrm{sl}_1$ into \eqref{solid_potential_eqs_b}, and assuming time-harmonic fields \eqref{time-harmonic}, we eliminate the temperature field and obtain a bi-harmonic equation for the scalar potential $\phi$,
 \bsub
 \bal
 \eqlab{phi_eq_1}
 \alpha_\mathrm{xl} \Lapl\Lapl\phi + \beta_\mathrm{xl} k_0^{2} \Lapl\phi + k_0^4\phi = 0,
 \text{ with } k_0 = \frac{\omega}{c},
 \eal
where we have introduced the undamped adiabatic wavenumber $k_0=\omega/c$, and where the parameters $\alpha_\mathrm{xl}$ and $\beta_\mathrm{xl}$ for fluids (xl = fl) and solids (xl = sl) are
\begin{alignat}{2}
\alpha_\mathrm{fl}  &= - \ii (1 - \ii \gamma \Gamv) \Gamt  , \quad
& \beta_\mathrm{fl} &= 1-\ii (\Gams + \gamma \Gamt) , \\
\alpha_\mathrm{sl}  &= - \ii (1+X) \Gamt , \quad
& \beta_\mathrm{sl} &= 1 - \ii \gamma \Gamt .
\end{alignat}
Here, we have used the relation \eqrefnoEq{c_ad_solid} for solids and further introduced the parameters $X$ and $\chi$,
\bal
\eqlab{X_def}
X &= (\gamma-1)(1-\chi), \\
\eqlab{chi_def}
\chi &= \dfrac{1}{\rho_0 \kapS c^2} = 1 - \dfrac{4}{3} \dfrac{\cT^2}{c^2},
\eal
\esub
the latter equality following from combining \eqref{c_ad_solid} with \eqref{kapT_of_cL_cT} and using $\kapT=\gamma\kapS$ from \eqref{gamma_small}. Note that for fluids $\chi = 1$, $\cT = 0$, and $X = 0$.

The bi-harmonic equation~\eqrefnoEq{phi_eq_1} is factorized and written on the equivalent form
\bsub
\bal
\eqlab{biharmonic_separated}
( \Lapl + \kc^2 ) ( \Lapl + \kt^2 ) \phi &= 0 ,
\eal
and thus the wavenumbers $\kc$ and $\kt$ are obtained from $\kc^2 + \kt^2 = \beta_\mathrm{xl} k_0^2/\alpha_\mathrm{xl}$ and $\kc^2\kt^2 = k_0^4/\alpha_\mathrm{xl}$, resulting in
\eqlab{kc_kT_exact}
\bal
\kc^2 &= 2 k_0^2 \left[ \beta_\mathrm{xl} + (\beta_\mathrm{xl}^2 - 4 \alpha_\mathrm{xl})^{1/2} \right]^{-1} , \\
\kt^2 &= 2 k_0^2 \left[ \beta_\mathrm{xl} - (\beta_\mathrm{xl}^2 - 4 \alpha_\mathrm{xl})^{1/2} \right]^{-1},
\eal
\esub
with "xl" being either "fl" for fluids or "sl" for solids.

In the frequency domain, the equation for the vector potential $\vpsi$, \eqref{first_order_eq_3_c} for fluids and \eqref{solid_potential_eqs_c} for solids, can be written as $\Lapl \vpsi + \ks^2\vpsi = \zerovec$, which describes a transverse shear mode with shear wavenumber $\ks$. By introducing a shear constant $\etaO$, which for a fluid is the dynamic viscosity, and for a solid is defined as,
 \bsub
 \beq{shear_constant}
 \eta_0 = \ii \: \dfrac{\rho_0 \cT^2}{\omega} \qquad \text{(solid)},
 \eeq
the shear wavenumber $\ks$ is given by the same expression for both fluids and solids,
 \bal
 \eqlab{ks_flsl_general}
 \ks^2 = \dfrac{\ii \omega \rho_0}{\eta_0} \qquad \text{(fluid and solid)}.
 \eal
 \esub

\subsection{Wave equations and modes}
The general solution $\phi$ of the bi-harmonic equation~\eqrefnoEq{biharmonic_separated} is the sum
\beq{phitot}
\phi = \phic + \phit
\eeq
of the two potentials $\phic$ and $\phit$, which satisfy the harmonic equations
 \bsub
 \eqlab{wave_eqs}
 \bal
 \Lapl \phic + \kc^2 \phic &= 0 , \eqlab{wave_eqs_a}\\
 \Lapl \phit + \kt^2 \phit &= 0 , \eqlab{wave_eqs_b}
 \eal
where $\phic$ describes a compressional propagating mode with wavenumber $\kc$, while $\phit$ describes a thermal mode with wavenumber $\kt$. These two scalar wave equations together with the vector wave equation for $\vpsi$, describing the shear mode with wavenumber $\ks$,
 \bal
 \Lapl \vpsi + \ks^2 \vpsi &= \vec{0}, \eqlab{wave_eqs_c}
 \eal
 \esub
comprise the full set of first-order equations in potential theory. These wave equations, coupled through the boundary conditions, govern acoustics in thermoviscous fluids and thermoelastic solids. The distinction between fluids and solids is to be found solely in the wavenumbers of the three modes.

\subsubsection{Approximate wavenumbers for fluids}
For most systems of interest, $\Gamv , \Gamt \ll 1$ allowing a simplification of the expressions for $\kc$ and $\kt$ in \eqref{kc_kT_exact}. To first order in $\Gamv$ and $\Gamt$ one finds
\bsub
\eqlab{wave_numbers}
\bal
\kc &= \dfrac{\omega}{c} \left[ 1 + \dfrac{\ii}{2}\left[\Gamv + (\gamma-1) \Gamt \right] \right] , \eqlab{wave_numbers_a} \\
\kt &= \dfrac{(1+\ii)}{\delt} \left[ 1 + \dfrac{\ii}{2}(\gamma-1)(\Gamv - \Gamt) \right], \eqlab{wave_numbers_b} \\
\ks &= \dfrac{(1+\ii)}{\dels} , \eqlab{wave_numbers_c}
\eal
\esub
where we have introduced the thermal diffusion length $\delt$ and the momentum diffusion length $\dels$. Heat and momentum diffuses from boundaries, such that the  characteristic thicknesses of the thermal and viscous boundary layers are $\delt$ and $\dels$, respectively, given by
 \beq{deltdels}
 \delt = \sqrt{\dfrac{2 \Dth}{\omega}} , \qquad
 \dels = \sqrt{\dfrac{2 \nu_0}{\omega}}.
 \eeq
For water at room temperature and 2 MHz frequency, $\dels \simeq 0.4~\SImum$, $\delt \simeq 0.2~\SImum$, and $\lambda \simeq 760~\SImum$. Consequently, the length scales of the thermal and viscous boundary layer thicknesses are the same order of magnitude and much smaller than the acoustic wavelength. With $k_0=\omega/c$ we note that
\bal
\Gamv = \dfrac{1}{2}(1+\beta)(k_0 \dels)^2 , \qquad \Gamt = \dfrac{1}{2} (k_0 \delt)^2 ,
\eal
and consequently
\bsub
\eqlab{wavenumbers_relative_magnitude}
\bal
\Gamv &\sim (k_0\dels)^2 \sim \Big|\frac{\kc}{\ks}\Big|^2 \ll 1, \\
\Gamt &\sim (k_0\delt)^2 \sim \Big|\frac{\kc}{\kt}\Big|^2  \ll 1.
\eal
\esub
In the long-wavelength limit of the scattering theory to be developed, we expand to first order in $k_0\dels$ and $k_0\delt$, and thus neglect the second-order quantities $\Gamv$ and $\Gamt$. For water at room temperature and MHz frequency one finds $k_0\dels \sim k_0\delt \sim 10^{-3}$, and $\Gamv \sim \Gamt \sim 10^{-6}$.

Clearly, the compressional mode with wavenumber $\kc$ describes a weakly damped propagating wave with $\im[\kc] \ll \re[\kc] \simeq \omega/c$. In contrast, $\im[\kt] \simeq \re[\kt]$ for the thermal mode and $\im[\ks] = \re[\ks]$ for the shear mode, which correspond to waves that are damped within their respective wavelengths. Hence, these modes describe boundary layers near interfaces of walls and particles, which decay exponentially away from these interfaces on the length scales set by $\delt$ and $\dels$.

\subsubsection{Approximate wavenumbers for solids}
Similar to the fluid case, we use the smallness of the thermal damping factor, $\Gamt\ll 1$, to expand the exact wavenumbers of \eqref{kc_kT_exact}. To first order we obtain
 \bsub
 \eqlab{wave_numbers_sl}
 \bal
 \eqlab{wave_numbers_sl_a}
 \kc &= \dfrac{\omega}{c} \left[ 1 + \dfrac{\ii}{2} (\gamma - 1) \chi \Gamt \right] ,
 \\
 \eqlab{wave_numbers_sl_b}
 \kt &= \dfrac{(1+\ii)}{\delt} \dfrac{1}{\sqrt{1-X}}\left[ 1 + \dfrac{\ii}{8} \dfrac{\gamma^2 \Gamt}{(1-X)} \right] , \\
 \eqlab{wave_numbers_sl_c}
 \ks &= \dfrac{\omega}{\cT} .
 \eal
 \esub
An important distinction between a fluid and a solid is that a solid allows for propagating transverse waves while a fluid does not. This is evident from the shear mode wavenumber $\ks$, which for solids is purely real, $\ks = \omega/\cT$, while for fluids $\im[\ks] = \re[\ks] = 1/\dels$.

\subsection{Acoustic fields from potentials}
For a given thermoacoustic problem, the boundary conditions are imposed on the acoustic fields  $\vvvI$, $\TI$, and $\sigmabfI$ and not directly on the potentials $\phic$, $\phit$, and $\vpsi$. We therefore need expressions for the acoustic fields in terms of the potentials in order to derive the boundary conditions for the latter.

The velocity fields follow trivially from the Helmholtz decompositions and are obtained from the same expression in both fluids and solids
 \bal
 \eqlab{velocity_field_of_phi}
 \vvvI  & =  \nablabf (\phic + \phit) + \rot\vpsi,\\ \nn
 \text{where }\; \vvvI &= - \ii \omega\uuuI \text{ for solids}.
 \eal
A single expression for $\TI$  in terms of $\phic$ and $\phit$, valid for both fluids and solids, is obtained from \eqref{T1_of_potential} in combination with Eqs.~\eqrefnoEq{kc_kT_exact} - \eqrefnoEq{wave_eqs} by introducing the material-dependent parameters $\bc$ and $\bt$,
 \bsub
 \eqlab{T1_bc_bt}
 \bal
 T_1 &= \bc \phic + \bt \phit,\\
 \eqlab{bc_coeff}
 \bc &= \dfrac{\ii \omega (\gamma - 1)}{\alpha_p c^2} , \quad
 \bt = \dfrac{1}{\chi \alpha_p \Dth}.
 \eal
 \esub
Here, we have neglected $\Gamv$ and $\Gamt$ relative to unity. Note that the ratio $\bc /\bt \sim \Gamt \ll 1$.

In a fluid, the pressure field $p_1$ is obtained by inserting \eqref{Helmholtz_decomp} into the momentum equation \eqrefnoEq{first_NS} and using the wave equations \eqrefnoEq{wave_eqs},
 \beq{p1_from_Phi}
 p_1 = \ii \omega \rho_0 (\phic+\phit) - (1+\beta) \eta_0 (\kc^2\phic + \kt^2\phit).
 \eeq
Inserting this expression into \eqref{visc_stress_tensor}, the stress tensor for fluids becomes,
\bal
\eqlab{stress_field_fl}
\vec{\sigma}_1 &= \eta_0 \left[ (2 \kc^2 - \ks^2) \phic + (2\kt^2 - \ks^2) \phit \right] \mathbf{I} \nn \\
&\quad + \eta_0 \left[ \nablabf \vvvI + (\nablabf \vvvI)^\mathrm{T} \right],
\eal
where $\vvvI$ can be expressed by the potentials through \eqref{velocity_field_of_phi}. This expression also holds true for the solid stress tensor \eqref{stress_tensor_sl} using the shear constant $\etaO$, \eqref{shear_constant}, and the velocity field $\vvvI = -\ii \omega\uuuI$, \eqref{dtu1}. This conclusion is obtained by inserting \eqref{T1_of_potential_solid} for $T_1^\mathrm{sl}$ into \eqref{stress_tensor_sl} for $\sigmabfI$ and using the wave equations~\eqrefnoEq{wave_eqs}.

\begin{figure*}[!!t]
\centering
\includegraphics[width=0.75\textwidth]{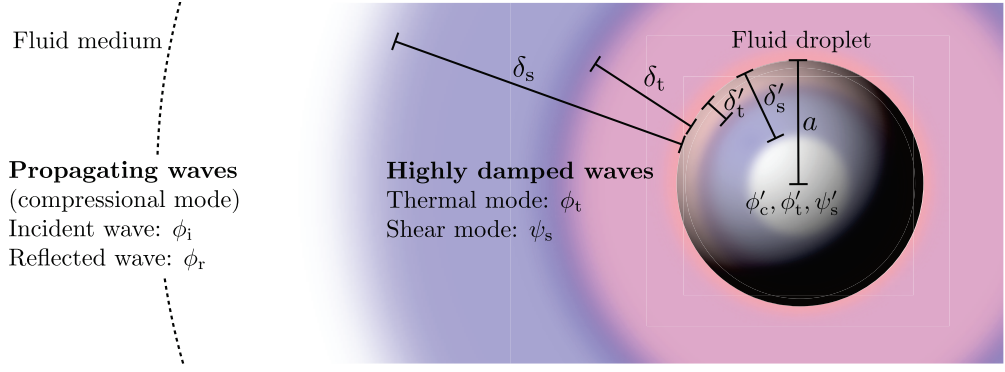}
\caption[]{(Color online) A compressional wave $\phi_\mathrm{i}$ propagating in a thermoviscous fluid medium with parameters $\rho_0,\eta_0,\kappa_s,\alpha_p,\cp,\gamma,$ and $\kth$, is incident on a thermoviscous fluid droplet with parameters $\rho_0^{\prime},\eta_0^{\prime},\kappa_s^{\prime},\alpha_p^{\prime},\cp^{\prime},\gamma^{\prime}$, and $\kth^{\prime}$, which results in a compressional scattered wave and highly damped thermal and shear waves  both outside in the fluid medium ($\phi_\mathrm{r},\phi_\mathrm{t},\psi_\mathrm{s}$) and inside in the fluid droplet ($\phi_\mathrm{c}^{\prime},\phi_\mathrm{t}^{\prime},\psi_\mathrm{s}^{\prime}$). Viscous and thermal boundary layers are described by the highly damped waves both outside and inside the fluid droplet. In the long-wavelength limit the droplet radius $a$ and the boundary layer thicknesses $\dels,\delt,\dels^{\prime},\delt^{\prime}$ are mutually unrestricted, but all much smaller than the acoustic wavelength $\lambda$. For a thermoelastic particle, the shear mode $\psi_s^\prime$ describes a propagating transverse wave instead of an internal viscous boundary layer.}
\figlab{rad_force_system}
\end{figure*}

\section{Scattering from a sphere}
\seclab{scattering_from_a_sphere}

The potential theory allows us in a unified manner to treat linear scattering of an acoustic wave on a spherical particle, consisting of either a thermoelastic solid or a thermoviscous fluid. The system of equations describing the general case of an arbitrary particle size is given, and analytical solutions are provided in the long-wavelength limit  $a , \dels , \delt \ll \lambda$. In this limit, the particle and boundary layers are much smaller than the acoustic wavelength, but the ratios $\dels/a$ and $\delt/a$ are unrestricted. This is essential for applying our results to micro- and nanoparticle acoustophoresis. In particular, we derive analytical expressions for the monopole and dipole scattering coefficients $f_0$ and $f_1$, which together with the incident acoustic field serve to calculate the acoustic radiation force as shown in \secref{Frad1} and summarized in \tabref{eqs_overview}.

\subsection{System setup}

We place the spherical particle of radius $a$ at the center of the coordinate system and use spherical coordinates $(r,\theta,\varphi)$ with the radial distance $r$, the polar angle $\theta$, and the azimuthal angle $\varphi$.  We let unprimed variables and parameters characterize the region of the fluid medium, $r > a$, while primed variables and parameters characterize the region of the particle, $r < a$. For example, the parameter $\kappa_s^{\prime}$ is the compressibility of the particle, while $\kappa_s$ is the compressibility of the fluid medium. Ratios of particle and fluid parameters are denoted by a tilde, e.g. $\tilde{\kappa}_s = \kappa_s^{\prime} / \kappa_s$. Due to linearity, we can without loss of generality assume that in the vicinity of the particle, the incident wave is a plane wave propagating in the positive $z$-direction, $\phi_i = \phi_0 \: \mathrm{e}^{\ii \kc z} = \phi_0 \: \mathrm{e}^{\ii \kc r \cos\theta}$. The fields do not depend on $\varphi$ due to azimuthal symmetry.

\subsection{Partial wave expansion}

The solution to the scalar and the vector wave equations \eqref{wave_eqs} with wavenumbers $k$, \eqsref{wave_numbers}{wave_numbers_sl},  in spherical coordinates is standard textbook material. Avoiding singular solutions at $r=0$ and considering outgoing scattered waves, the solution is written in terms of spherical Bessel functions $j_n(k r)$, outgoing spherical Hankel functions $h_n(k r)$, and Legendre polynomials $P_n(\cos\theta)$. As a consequence of azimuthal symmetry, only the $\varphi$-component of the vector potential is non-zero, $\vpsi(\rrr) = \psis(r,\theta) \: \textbf{e}_{\varphi}$. The solution is written as a partial wave expansion of the incident propagating wave $\phi_\mathrm{i}$, the scattered reflected propagating wave $\phi_\mathrm{r}$, the scattered thermal wave $\phit$, and the scattered shear wave $\psis$:\\

\noindent
\underline{In the fluid medium, $r > a$}
\bsub
 \eqlab{partial_wave_expansion}
 \bal
 \eqlab{phiInDef}
 \phi_\mathrm{i} &= \phi_0 \sum_{n=0}^{\infty} \ii^n (2n+1) j_n(\kc r) P_n(\cos\theta) , \\
 \phi_\mathrm{r} &= \phi_0 \sum_{n=0}^{\infty} \ii^n (2n+1) A_n h_n(\kc r) P_n(\cos\theta) ,\\
 \phit &= \phi_0 \sum_{n=0}^{\infty} \ii^n (2n+1) B_n h_n(\kt r) P_n(\cos\theta) , \\
 \psis &= \phi_0 \sum_{n=0}^{\infty} \ii^n (2n+1) C_n h_n(\ks r) \pp_\theta P_n(\cos\theta),
 \eal
\noindent \underline{In the particle, $r < a$}
 \bal
 \phic^{\prime} &= \phi_0 \sum_{n=0}^{\infty} \ii^n (2n+1) A_n^{\prime} j_n(\kc^{\prime} r) P_n(\cos\theta) , \\
 \phit^{\prime} &= \phi_0 \sum_{n=0}^{\infty} \ii^n (2n+1) B_n^{\prime} j_n(\kt^{\prime} r) P_n(\cos\theta) , \\
 \psis^{\prime} &= \phi_0 \sum_{n=0}^{\infty} \ii^n (2n+1) C_n^{\prime} j_n(\ks^{\prime} r) \pp_\theta P_n(\cos\theta),
 \eal
\esub
where the parameter $\phi_0$ is an arbitrary amplitude of the incident wave with unit m$^2\:$s$^{-1}$. The different components of the resulting acoustic field are illustrated in \figref{rad_force_system}.

\subsection{Boundary conditions}
Neglecting surface tension, the appropriate boundary conditions at the particle surface are continuity of velocity, normal stress, temperature, and heat flux. Assuming sufficiently small oscillations, see \secref{OscParticle}, the boundary conditions are imposed at $r=a$,
\bsub
\eqlab{scattering_bc}
\begin{alignat}{3}
v_{1r} &= v_{1r}^{\prime},               &   v_{1\theta} &= v_{1\theta}^{\prime}, & T_1 &= T_1^{\prime} , \\[2mm]
\sigma_{1r r} &= \sigma_{1r r}^{\prime},
\quad&  \sigma_{1 \theta r} &= \sigma_{1 \theta r}^{\prime},
\quad& \kth \pp_r T_1 &= \kth^{\prime} \pp_r T_1^{\prime}.
\end{alignat}
\esub
The boundary conditions are expressed in terms of the potentials using \eqssref{velocity_field_of_phi}{T1_bc_bt}{stress_field_fl}. The components of velocity and stress in spherical coordinates are given in \appref{stress_tensor_app}.

It is convenient to introduce the non-dimensionalized wavenumbers $\xc$, $\xt$, and $\xs$ for the medium, and $\xc^\prime$, $\xt^\prime$, and $\xs^\prime$ for the particle,
\bsub
\eqlab{xxDef}
\begin{alignat}{3}
\xc &= \kc a , & \qquad \xt &= \kt a ,& \qquad \xs &= \ks a , \\
\xc^\prime &= \kc^\prime a ,&  \xt^\prime &= \kt^\prime a,& \xs^\prime &= \ks^\prime a .
\end{alignat}
\esub
Inserting the expansion \eqrefnoEq{partial_wave_expansion} into the boundary conditions \eqrefnoEq{scattering_bc}, and making use of the Legendre equation \eqrefnoEq{legendre_eq}, we obtain the following system of coupled linear equations for the expansion coefficients in each order $n$,\\

\noindent $\underline{a v_{1 r} = a v_{1 r}^\prime}$
\bsub
\eqlab{explicit_bc_eqs_general}
 \begin{align}
 & \xc j_n^{\prime}(\xc) \!+\! A_n \xc h_n^{\prime}(\xc) \!+\! B_n \xt h_n^{\prime}(\xt) \!-\! C_n n(n\!+\!1)  h_n(\xs) \nn \\
 & = A_n^{\prime} \xcp j_n^{\prime}(\xcp) + B_n^{\prime} \xtp j_n^{\prime}(\xtp) - C_n^{\prime} n(n\!+\!1)  j_n(\xsp),
 \eqlab{bc_eqs_a}
 \end{align}

\noindent
$\underline{a v_{1 \theta} = a v_{1 \theta}^\prime}$
\begin{align}
& j_n(\xc) + A_n h_n(\xc) + B_n h_n(\xt) - C_n \big[ \xs h_n^{\prime}(\xs) + h_n(\xs) \big] \nn \\
& = A_n^{\prime} j_n(\xcp) +\! B_n^{\prime} j_n(\xtp) -\! C_n^{\prime} \big[\xsp j_n^{\prime}(\xsp) + j_n(\xsp) \big],
\eqlab{bc_eqs_b}
\end{align}
$\underline{T_1 = T_1^\prime}$
\begin{align}
& \bc j_n(\xc) + A_n \bc h_n(\xc) + B_n \bt h_n(\xt) \nn \\
& = A_n^{\prime} \bcp j_n(\xcp) + B_n^{\prime} \btp j_n(\xtp) ,
\eqlab{bc_eqs_c}
\end{align}
$\underline{a \kth \pp_r T_1 = a \kth^\prime \pp_r T_1^\prime}$
\begin{align}
&\kth \bc \xc j_n^{\prime}(\xc) + A_n \kth \bc \xc h_n^{\prime}(\xc) + B_n \kth \bt \xt h_n^{\prime}(\xt) \nn \\
& = A_n^{\prime} \kth^{\prime} \bcp \xcp j_n^{\prime}(\xcp) + B_n^{\prime} \kth^{\prime} \btp \xtp  j_n^{\prime}(\xtp) ,
\eqlab{bc_eqs_d}
\end{align}
$\underline{a^2 \sigma_{1 \theta r} = a^2 \sigma_{1 \theta r}^\prime}$
\begin{align}
& \eta_0 \left[ \xc j_n^{\prime}(\xc) - j_n(\xc) \right]+ A_n \eta_0 \left[ \xc h_n^{\prime}(\xc)-h_n(\xc) \right] \nn \\
& \qquad + B_n \eta_0 \left[ \xt h_n^{\prime}(\xt) - h_n(\xt) \right]  \nn \\
& \qquad - \frac12 C_n \eta_0 \left[ \xs^2 h_n^{\prime\prime}(\xs) + (n^2+n-2) h_n(\xs) \right] \nn \\
& = A_n^{\prime} \eta_0^{\prime} \left[ \xcp j_n^{\prime}(\xcp)-j_n(\xcp) \right] + B_n^{\prime} \eta_0^{\prime} \left[ \xtp j_n^{\prime}(\xtp) - j_n(\xtp) \right] \nn \\
& \qquad - \frac12 C_n^{\prime} \eta_0^{\prime} \left[ \xs^{\prime 2} j_n^{\prime\prime}(\xsp) + (n^2+n-2) j_n(\xsp) \right] ,
\eqlab{bc_eqs_e}
\end{align}
$\underline{a^2 \sigma_{1 r r} = a^2 \sigma_{1 r r}^\prime}$
\begin{align}
& \eta_0 \left[ (\xs^2 - 2 \xc^2 ) j_n (\xc) - 2 \xc^2 j_n^{\prime\prime}(\xc) \right] \nn \\
& \qquad + A_n \eta_0 \left[ (\xs^2 - 2 \xc^2) h_n(\xc) - 2 \xc^2 h_n^{\prime\prime}(\xc) \right] \nn \\
& \qquad + B_n \eta_0 \left[ (\xs^2 - 2 \xt^2) h_n(\xt) - 2 \xt^2 h_n^{\prime\prime}(\xt) \right] \nn \\
& \qquad + 2n(n+1) C_n \eta_0 \left[ \xs h_n^{\prime}(\xs) - h_n(\xs) \right] \nn \\
& = A_n^{\prime} \eta_0^{\prime} \left[ (\xs^{\prime 2} - 2 \xc^{\prime 2}) j_n(\xcp) - 2 \xc^{\prime 2} j_n^{\prime\prime}(\xcp) \right] \nn \\
& \qquad + B_n^{\prime} \eta_0^{\prime} \left[ (\xs^{\prime 2} - 2 \xt^{\prime 2}) j_n(\xtp) - 2 \xt^{\prime 2} j_n^{\prime\prime}(\xtp) \right] \nn \\
& \qquad + 2n(n+1) C_n^{\prime} \eta_0^{\prime} \left[ \xsp j_n^{\prime}(\xsp) - j_n(\xsp) \right] .
\eqlab{bc_eqs_f}
\end{align}
\esub
Here, primes on spherical Bessel and Hankel functions indicate derivatives with respect to the argument. The equations are valid for both a fluid and solid particle, with $\eta_0^{\prime}$ being the viscosity for a fluid particle and the shear constant \eqref{shear_constant} for a solid particle.

For $n=0$, the boundary conditions for $v_{1\theta}$ and $\sigma_{1 \theta r}$ are trivially satisfied because there is no angular dependence in the zeroth-order Legendre polynomial, $P_0(\cos\theta)=1$. Consequently, $\psis=0$, and we are left with four equations with four unknowns, namely Eqs.~\eqrefnoEq{bc_eqs_a}, \eqrefnoEq{bc_eqs_c}, \eqrefnoEq{bc_eqs_d}, and \eqrefnoEq{bc_eqs_f} with $C_0=C_0^\prime=0$.

The linear system of equations \eqrefnoEq{explicit_bc_eqs_general} may be solved for each order $n$ yielding the scattered field with increasing accuracy as higher-order multipoles are taken into account, an approach referred to within the field of ultrasound characterization of emulsions and suspensions as ECAH theory after Epstein and Carhart \cite{Epstein1953} and Allegra and Hawley \cite{Allegra1972}. However, care must be taken due to the system matrix often being ill-conditioned~\cite{Pinfield2007}.

The long-wavelength limit is characterized by the small dimensionless parameter $\ve$, given by
 \beq{eps_def}
 \ve = k_0 a = 2\pi\:\frac{a}{\lambda} \ll 1,
 \eeq
In this limit, the dominant contributions to the scattered field are due to the $n=0$ monopole and the $n=1$ dipole terms, both proportional to $\ve^3$, while the contribution of the  $n$th-order multipole for $n>1$ is proportional to $\ve^{2n+1} \ll \ve^3$.

\subsection{Monopole scattering coefficient}
To obtain the monopole scattering coefficient $f_0$ in \eqref{Frad_Settnes}, we solve for the expansion coefficient $A_0$ in \eqref{explicit_bc_eqs_general} and use the identity $f_0 = 3\ii\: \xc^{-3} A_0$.
The $f_n$-coefficients are traditionally used in work on the acoustic radiation force, while the $A_n$-coefficients are used in general scattering theory.

The solution to the inhomogeneous system of linear equations for $n=0$ involves straightforward but lengthy algebra presented in \appref{monopole_calculations}. In \eqref{f0_general_app} is given the general analytical expression for $f_0$ in the long-wavelength limit valid for any particle. In the following, this expression is given in explicit, simplified, closed analytical form for a thermoviscous droplet and a thermoelastic particle, respectively.

\subsubsection{A thermoviscous droplet in a fluid}
\seclab{droplet_fluid}
For a thermoviscous droplet in a fluid in the long-wavelength limit, the particle radius $a$ and the viscous and thermal boundary layers both inside ($\dels'$, $\delt'$) and outside  ($\dels$, $\delt$) the fluid droplet are all much smaller than the acoustic wavelength $\lambda$, while nothing is assumed about the relative magnitudes of
$a$, $\dels$, $\dels'$, $\delt$, and $\delt'$.  Thus, using the non-dimensionalized wavenumbers \eqref{xxDef} and $\ve = \kO a$, the long-wavelength limit is defined as
\bsub
\eqlab{fluid-fluid_longwave}
\bal
|\xc|^2 , |\xc^{\prime}|^2  \sim \ve^2 &\ll 1 \quad\text{and}\quad
\\
|\xc|^2 , |\xc^{\prime}|^2  \sim \ve^2 &\ll|\xs|^2 , |\xs^\prime|^2 , |\xt|^2 , |\xt^{\prime}|^2 ,
\eal
which implies
\bal
\Gams , \Gamt , \frac{|\bc|}{|\bt|} , \frac{|\bcp|}{|\bt|} \sim  \ve^2 \ll 1 .
\eal
\esub
To first order in $\ve$, the analytical result for the monopole scattering coefficient $\fOfl$ obtained from  \eqref{f0_general_app} is most conveniently written as
 \bsub
 \eqlab{f0_fluid-fluid}
 \bal
 &\fOfl = 1 - \tilde{\kappa}_s + 3 (\gamma - 1 )
 \left(1 - \dfrac{ \alfpTi }{\rhoOTi \: \tilde{c}_p} \right)^2 H(\xt,\xtp),
 \\
 \eqlab{monopole_H}
 &H(\xt,\xtp) = \dfrac{1}{\xt^2} \left[ \dfrac{1}{1-\ii \xt} - \dfrac{1}{\tilde{k}_\mathrm{th}} \dfrac{\tan \xt^{\prime}}{\tan \xt^{\prime} - \xt^{\prime}} \right]^{-1},
 \eal
 \esub
where $H(\xt,\xtp)$ is a function of the particle radius $a$ through the non-dimensionalized thermal wavenumbers $\xt$ and $\xt^\prime$. Epstein and Carhart obtained a corresponding result for $A_0$ but with a sign-error in the thermal correction term \cite{Epstein1953}, while the result of Allegra and Hawley \cite{Allegra1972} is in agreement with what we present here. The factor $(\gamma-1)$ quantifies the coupling between heat and the mechanical pressure waves. This factor is multiplied by $\big[1 - \alfpTi / (\rhoOTi \tilde{c}_p)\big]^2$, where the quantity $\xi_p = \alpha_p / (\rho_0 \cp)$, with unit m$^3/$J, may be interpreted as an isobaric expansion coefficient per added heat unit. The thermal correction can only be non-zero if there is a contrast $\tilde{\xi}_p \neq 1$ in this parameter.

In the weak dissipative limit of small boundary layers the function $H(\xt,\xtp)$ is expanded to first order in $\delt/a$ and $\delt'/a$, and using $\tan(\xt') \simeq \ii$, we obtain
 \bsub
 \eqlab{f0_fluid-fluid_weak}
 \bal
 &\fOfl = 1 - \tilde{\kappa}_s - \dfrac{3}{2}\dfrac{(1+\ii)(\gamma-1)}{1+\tilde{D}_\mathrm{th}^{1/2} \tilde{k}_\mathrm{th}^{-1}} \left( 1 - \dfrac{\alfpTi}{\rhoOTi \tilde{c}_p} \right)^2 \dfrac{\delt}{a},
 \\ &\nn \hspace*{10em} \text{(Small-width boundary layers)},
 \\
 &|\xc|^2 , |\xc^{\prime}|^2 \ll 1 \ll
 |\xs|^2 , |\xs^\prime|^2 , |\xt|^2 , |\xt^{\prime}|^2.
 \eal
 \esub
In the limit of zero boundary-layer thickness $\delt/a \rightarrow 0$, the thermal correction vanishes, and we obtain
 \bsub
 \eqlab{f0_fluid-fluid_large}
 \bal
 &\fOfl = 1 - \tilde{\kappa}_s, \quad\quad \text{(Zero-width boundary layers),}
 \\
 \eqlab{f0_fluid-fluid_large_limitDef}
 &|\xc|^2 , |\xc^{\prime}|^2 \ll 1 \text{ and }
 |\xs|^2 , |\xs^\prime|^2 , |\xt|^2 , |\xt^{\prime}|^2 \rightarrow \infty ,
 \eal
 \esub
which is the well-known result for a compressible sphere in an ideal \cite{Gorkov1962} or a viscous \cite{Settnes2012} fluid.

In the opposite limit of a point particle, $a/\delt,a/\delt' \rightarrow 0$, we find $H(\xt,\xtp) = -(1/3)\rhoOTi \tilde{c}_p$, yielding
 \bsub
 \eqlab{f0_fluid-fluid_point}
 \bal
 &\fOfl = 1 - \tilde{\kappa}_s - (\gamma - 1 ) \rhoOTi \tilde{c}_p \left(1 - \dfrac{ \alfpTi }{\rhoOTi \: \tilde{c}_p} \right)^2,
 \\
 &\nn \hspace*{14em} \text{(Point-particle limit)},
 \\
 \eqlab{f0_fluid-fluid_point_limitDef}
 &|\xc|^2 , |\xc^{\prime}|^2 \ll 1 \text { and }
 |\xs|^2 , |\xs^\prime|^2 , |\xt|^2 , |\xt^{\prime}|^2 \rightarrow 0.
 \eal
 \esub
Since $\gamma>1$, the correction from thermal effects in the point-particle limit is negative. This implies that the thermal correction enhances the magnitude of $\fOfl$ for acoustically soft particles  ($\kapsTi > 1$), while it diminishes the magnitude and eventually may reverse the sign of $\fOfl$ for acoustically hard particles ($\kapsTi < 1$).

Importantly, an inspection of the point-particle limit \eqref{f0_fluid-fluid_point} leads to two noteworthy conclusions not previously discussed in the literature. Firstly, the thermal contribution to $\fOfl$ allows for a sign change of the acoustic radiation force for different-sized but otherwise identical particles. Secondly, the thermal contribution may result in forces orders of magnitude larger than expected from both ideal \cite{Gorkov1962} and viscous \cite{Settnes2012} theory. For example, $\rhoOTi \gg 1$ for particles or droplets in gases leads to a thermal contribution to $\fOfl$ two orders of magnitude larger than $1-\kapsTi$. These predictions are discussed in more detail in \secref{examples}.

\subsubsection{A thermoelastic particle in a fluid}
\seclab{particle_fluid}
For a thermoelastic particle in a fluid, the long-wavelength limit differs from that of a thermoviscous droplet \eqref{fluid-fluid_longwave} by the shear mode describing a propagating wave and not a viscous boundary layer.
The wavelength of this transverse shear wave is comparable to that of the longitudinal compressional wave, and in the long-wavelength limit both are assumed to be large,
\bsub
\eqlab{particle-fluid_longwave}
\bal
|\xc|^2 , |\xc^{\prime}|^2, |\xs^\prime|^2  \sim \ve^2 &\ll 1 \quad\text{and}\quad
\\
|\xc|^2 , |\xc^{\prime}|^2, |\xs^\prime|^2   \sim \ve^2 &\ll|\xs|^2 , |\xt|^2 , |\xt^{\prime}|^2 ,
\eal
which implies
\bal
\Gams , \Gamt ,   \frac{1}{|\etaOTi|}, \frac{|\bc|}{|\bt|} , \frac{|\bcp|}{|\bt|} \sim  \ve^2 \ll 1 .
\eal
\esub
To first order in $\ve$, the result \eqref{f0_general_app} for $\fOsl$ may be simplified as outlined in \appref{app_scattering_coeffs}, and one obtains after some manipulation
\begin{widetext}
\bal
\eqlab{f0_fluid-solid}
\fOsl &= \dfrac{1-\tilde{\kappa}_s + 3 (\gamma - 1) \left[ \left( 1 - \dfrac{\alfpTi}{\rhoOTi \tilde{c}_p} \right) \left( 1 - \dfrac{\chi^\prime \alfpTi }{\rhoOTi \tilde{c}_p} \right) - \dfrac{4}{3} \dfrac{\chi^\prime \alfpTi \tilde{\kappa}_s }{\tilde{c}_p} \dfrac{\cTpsqr}{c^2} \left( 1 - \dfrac{\alfpTi}{\rhoOTi \tilde{c}_p \tilde{\kappa}_s} \right)\right] H(\xt,\xtp) }{1 + 4 (\gamma - 1) \dfrac{\chi^\prime \alfpTi^2 }{\rhoOTi \tilde{c}_p^2} \dfrac{\cTpsqr}{c^2} H(\xt,\xtp)},
\eal
where the function $H(\xt,\xtp)$ is still given by the expression in \eqref{monopole_H} with $\xtp$ being the non-dimensionalized thermal wavenumber in the solid particle obtained from \eqref{wave_numbers_sl_b}. In the limit of a point particle, $a/\delt, a/\delt' \rightarrow 0$, we find
 \bsub
 \eqlab{f0_fluid-solid_point}
 \bal
 \fOsl &= \dfrac{1-\tilde{\kappa}_s - \dfrac{(\gamma - 1)\rhoOTi \tilde{c}_p }{1-X'}
 \left[ \left( 1 - \dfrac{\alfpTi}{\rhoOTi \tilde{c}_p} \right) \left( 1 - \dfrac{\chi^\prime \alfpTi }{\rhoOTi \tilde{c}_p} \right) - \dfrac{4}{3} \dfrac{\chi^\prime \alfpTi \tilde{\kappa}_s }{\tilde{c}_p} \dfrac{\cTpsqr}{c^2} \left( 1 - \dfrac{\alfpTi}{\rhoOTi \tilde{c}_p \tilde{\kappa}_s} \right)\right] }{1 - \dfrac{4}{3} \dfrac{\gamma - 1}{1-X'} \dfrac{\chi^\prime \alfpTi^2 }{\tilde{c}_p} \dfrac{\cTpsqr}{c^2} } \quad \text{(Point-particle limit)},\\
 \eqlab{f0_fluid-solid_point_limitDef}
 & \hspace*{26em} |\xc|^2 , |\xc^{\prime}|^2, |\xs^\prime|^2  \ll 1 \text { and }
 |\xs|^2 , |\xt|^2 , |\xt^{\prime}|^2 \rightarrow 0.
 \eal
 \esub
\end{widetext}
Remarkably, in the point-particle limit $\fOsl$ and $\fOfl$ differ in general. However, as expected, letting $\cTp \rightarrow 0$ in \eqref{f0_fluid-solid},  $\fOsl$ reduces to $\fOfl$ \eqref{f0_fluid-fluid} for all particle sizes.

In the weak dissipative limit of small boundary layers, $\delt,\delt^\prime\ll a$,
the second term in the denominator of \eqref{f0_fluid-solid} is small for typical material parameters. An expansion in $\delt/a$ and $\delt'/a$ then yields in analogy with \eqref{f0_fluid-fluid_weak},
 \bsub
 \eqlab{f0_fluid-solid_weak}
 \bal
 & \fOsl = 1 - \tilde{\kappa}_s - \dfrac{3}{2}\dfrac{(1+\ii)(\gamma-1)}{1+(1\!-\!X^\prime)^{1/2}\tilde{D}_\mathrm{th}^{1/2} \tilde{k}_\mathrm{th}^{-1}} \left(\! 1 - \dfrac{\alfpTi}{\rhoOTi \tilde{c}_p} \right)^2 \dfrac{\delt}{a}  %
 \\ &\nn \hspace*{10em} \text{(Small-width boundary layers)},
 \\
 & |\xc|^2 , |\xc^{\prime}|^2 , |\xs^\prime|^2  \ll 1 \ll
 |\xs|^2, |\xt|^2 , |\xt^{\prime}|^2,
 \eal
 \esub
simplified using \eqref{chi_def}. In the limit $\delt/a \rightarrow 0$, the thermal correction terms vanishes,
 \bsub
 \eqlab{f0_fluid-solid_large}
 \bal
 &\fOsl = 1 - \tilde{\kappa}_s,  \quad\quad   \text{(Zero-width boundary layers),}
 \\
 \eqlab{f0_fluid-solid_large_limitDef}
 &|\xc|^2 , |\xc^{\prime}|^2, |\xs^\prime|^2 \ll 1 \text{ and }
 |\xs|^2  , |\xt|^2 , |\xt^{\prime}|^2 \rightarrow \infty ,
 \eal
 \esub
In this limit, where boundary layer effects are negligible, $\fOsl$ and $\fOfl$ are identical and, as expected, equal to the ideal \cite{Gorkov1962} and viscous \cite{Settnes2012} results.

\subsection{Dipole scattering coefficient}
To obtain the dipole scattering coefficient $f_1$ in \eqref{Frad_Settnes}, we solve for the expansion coefficient $A_1$ in \eqref{explicit_bc_eqs_general} and use the identity $f_1 = -6\ii\: \xc^{-3} A_1$. In the long-wavelength limit, the terms involving the coefficients $B_1$ and $B_1^{\prime}$ are neglected to first order in $\ve$. This reduces the system of equations \eqrefnoEq{explicit_bc_eqs_general} for $n=1$ from six to four equations with the unknowns $A_1$, $A'_1$, $C_1$, and $C'_1$. In \appref{dipole_calculations} we solve explicitly for $A_1$. Physically, the smallness of the $B_1$- and $B_1^{\prime}$-terms means that thermal effects are negligible compared to viscous effects. This is consistent with the dipole mode describing the center-of-mass oscillations of the undeformed particle.

\subsubsection{A thermoviscous droplet in a fluid}
The analytical expression for $A_1$ in the long-wavelength limit for a thermoviscous droplet in a fluid, as defined in \eqref{fluid-fluid_longwave}, is given in \eqref{dipole_A1} of \appref{dipole_calculations}. This expression for $A_1$ was also obtained by Allegra and Hawley \cite{Allegra1972} and, with a minor misprint, by Epstein and Carhart \cite{Epstein1953} in their studies of sound attenuation in emulsions and suspensions. We write the result for the dipole scattering coefficient $f_1$ on a form more suitable for comparison to the theory of acoustic radiation forces as presented by Gorkov \cite{Gorkov1962} and Settnes and Bruus \cite{Settnes2012},
 \bsub
 \eqlab{f1_fluid-fluid}
 \bal
 \eqlab{f1_fl_final}
 \fIfl &= \dfrac{2\left(\tilde{\rho_0}-1\right)\left(1+F(\xs,\xsp) -G(\xs) \right)}
 {\left(2\tilde{\rho_0}+1\right)\big[1+F(\xs,\xsp)\big]-3 G(\xs)} ,
 \\[2mm]
 \eqlab{G_def}
 G(\xs) &= \dfrac{3}{\xs} \left(\dfrac{1}{\xs} - \ii \right),
 \\[2mm]
 \eqlab{F_def}
 F(\xs,\xsp) & = \dfrac{1 - \ii \xs}{2 (1- \etaOTi ) + \dfrac{\etaOTi \xs^{\prime 2} (\tan \xs^{\prime} - \xs^{\prime}
 )}{(3-\xs^{\prime 2})\tan \xs^{\prime} - 3 \xs^{\prime}}}.
  \eal
  \esub
Even though no thermal effects are present in $\fIfl $, \eqref{f1_fluid-fluid} is nevertheless an extension of the result by Settnes and Bruus \cite{Settnes2012}, since we have taken into account a finite viscosity in the droplet entering through the parameters $\etaOTi$ and $\xsp$. In the limit $\etaOTi \rightarrow \infty$ of infinite droplet viscosity, the function $F(\xs,\xsp)$ tends to zero, and we recover the result for $f_1$ obtained in Ref.~\cite{Settnes2012}.

In the weak dissipative limit of small boundary layers, $\dels,\dels^\prime \ll a$, the dipole scattering coefficient for the thermoviscous droplet reduces to
\bsub
\eqlab{f1_fluid-fluid_weak}
\bal
 &\fIfl = \dfrac{2(\rhoOTi-1)}{2\rhoOTi+1} \left[ 1 + \dfrac{3(1+\ii)}{1+\tilde{\nu}_0^{1/2} \etaOTi^{\:-1}} \dfrac{\rhoOTi-1}{2\rhoOTi + 1} \dfrac{\dels}{a} \right] ,
 \\ &\nn \hspace*{10em} \text{(Small-width boundary layers)},
 \\
 &|\xc|^2 , |\xc^{\prime}|^2 \ll 1 \ll
 |\xs|^2 , |\xs^\prime|^2 , |\xt|^2 , |\xt^{\prime}|^2.
 \eal
 \esub

\subsubsection{A thermoelastic particle in a fluid}
In the long-wavelength limit \eqref{particle-fluid_longwave} of a thermoelastic solid particle in a fluid, we obtain the result
\bal
\eqlab{f1_fluid-solid}
\fIsl = \dfrac{2\left(\tilde{\rho_0}-1\right)\left(1-G(\xs) \right)}{2\tilde{\rho_0}+1-3 G(\xs)},
\eal
with the function $G(\xs)$ given in \eqref{f1_fluid-fluid}. In this expression, the only particle-related parameters are density and radius, and it is identical to that derived by Settnes and Bruus \cite{Settnes2012}, who included the same two parameters in their study of scattering from a compressible particle in a viscous fluid using asymptotic matching.

In the small-width boundary layer limit, $\dels \ll a$, the dipole scattering coefficient for the thermoelastic solid particle $\fIsl$ reduces
\bsub
 \eqlab{f1_fluid-solid_weak}
 \bal
 &\fIsl = \dfrac{2(\rhoOTi-1)}{2\rhoOTi+1} \left[ 1 + 3(1+\ii) \dfrac{\rhoOTi-1}{2\rhoOTi +   1} \dfrac{\dels}{a} \right] ,
 \\ &\nn \hspace*{10em} \text{(Small-width boundary layers)},
 \\
 & |\xc|^2 , |\xc^{\prime}|^2 , |\xs^\prime|^2  \ll 1 \ll
 |\xs|^2, |\xt|^2 , |\xt^{\prime}|^2,
 \eal
 \esub
which closely resembles \eqref{f1_fluid-fluid_weak} for $\fIfl$.

\subsubsection{Asymptotic limits}
In the zero-width boundary layer limit, the dipole scattering coefficients $\fIfl$ and $\fIsl$ both reduce to the ideal-fluid expression \cite{Gorkov1962},
 \bal
 \eqlab{f1_large}
 \fIfl = \fIsl = \dfrac{2 (\tilde{\rho_0}-1)}{2\tilde{\rho_0}+1} ,
 \quad \text{(Zero-width boundary layers),}
 \eal
with the zero-width boundary layer limit defined for a droplet and a solid particle in \eqsref{f0_fluid-fluid_large_limitDef}{f0_fluid-solid_large_limitDef}, respectively.

In the opposite limit of a point particle, $F(\xs,\xsp)=1/(2+3\etaOTi)$ is finite and the expression for $\fOfl$ and $\fOsl$ is dominated by the $G(\xs)$ terms, with both cases yielding the asymptotic result
 \bal
 \eqlab{f1_point}
 \fIfl = \fIsl = \dfrac{2}{3} ( \rhoOTi - 1) ,
 \quad\quad \text{(Point-particle limit)} ,
 \eal
with the point-particle limit defined for a droplet and a solid particle in \eqsref{f0_fluid-fluid_point_limitDef}{f0_fluid-solid_point_limitDef}, respectively. It is remarkable that for small particles suspended in a gas, where $\rhoOTi \gg 1$, the value of $f_1$ in \eqref{f1_point} is three to five orders of magnitude larger than the value $f_1 = 1$ predicted by ideal-fluid theory \cite{Gorkov1962}.

\section{Range of validity}
\seclab{constraints}

Before turning to experimentally relevant predictions derived from our theory, we discuss the range of validity of our results imposed by the three main assumptions: the time periodicity of the total acoustic fields, the perturbation expansion of the acoustic fields, and the restrictions associated with size, shape and motion of the suspended particle.

\subsection{Time periodicity}

The first fundamental assumption in our theory is the restriction to time-periodic total acoustic fields, which was used to obtain \eqref{FradExactFF} for the acoustic radiation force evaluated at the static far-field surface $\pp\Omega_1$. Given a time-harmonic incident field, as studied in this work, a violation of time periodicity can only be caused by a non-zero time-averaged drift of the suspended particle. Denoting the speed of this drift by $\vp(t)$, we consider first the case of a steady particle drift. The assumption of time periodicity is then a good approximation if the displacement $\Delta \ell$ is small compared to the particle radius $a$ during one acoustic oscillation cycle $\tau = 2\pi/\omega$ used in the time averaging. A non-zero, acoustically-induced particle drift speed $\vp$ must be of second or higher order in $\veac$, $\vp /c \sim \veac^2$, as all first-order fields have a zero time average. Thus
 \beq{particle_drift}
 \frac{\Delta\ell}{a} \simeq \frac{\vp\tau}{a}
 = \frac{2\pi \vp}{\omega a}
 = \frac{2\pi}{k_0 a} \dfrac{\vp}{c}
 = \frac{\lambda}{a}\:\veac^2 \ll 1,
 \eeq
and time periodicity is approximately upheld for reasonably small perturbation strengths $\veac \ll \sqrt{a/\lambda}$, which is not a severe restriction in practice. In a given experimental situation, it is also easy to check if a measured non-zero drift velocity fulfills $\vp \tau \ll a$.

In the case of an unsteady drift speed $\vp(t)$, the time-averaged rate of change of momentum $\avr{\frac{\dm \PPP}{\dm t}}$ in the fluid volume bounded by $\pp\Omega_1$ in \eqref{PrateFlux1} is non-zero, thus violating the assumption $\avr{\frac{\dm \PPP}{\dm t}} = \vec{0}$ leading to \eqref{FradExactFF}. Only the unsteady growth of the viscous boundary layer in the fluid surrounding the accelerating particle contributes to $\avr{\frac{\dm \PPP}{\dm t}}$, since equal amounts of momentum is fluxed into and out of the static fluid volume in the steady problem. For \eqref{FradExactFF} to remain approximately valid, we must require $\avr{\frac{\dm \PPP}{\dm t}}$ to be much smaller than $\FFFrad$. To check this requirement, we consider a constant radiation force accelerating the particle. When including the added mass from the fluid, this leads to the well known time-scale $\taup$ for the acceleration,
 \beq{tau_ratio}
 \taup = \frac{2\rhoOTi + 1}{9\pi} \frac{a^2}{\dels^2}\:\tau.
 \eeq
Thus, small particles ($a\ll\dels$) are accelerated to their steady velocity in a timescale much shorter than the acoustic oscillation period ($\taup\ll\tau$), while the opposite ($\taup\gg\tau$) is the case for large particles ($a\gg\dels$). The unsteady momentum transfer to the fluid bounded by $\pp\Omega_1$ is obtained from the unsteady part $F^\mathrm{unst}_\mathrm{drag}(t)$ of the drag force on the particle as $\avr{\frac{\dm P}{\dm t}}=\frac{1}{\tau}\int_0^\tau F^\mathrm{unst}_\mathrm{drag}(t)\:\dm t$. Using the explicit expression for $F_\mathrm{drag}(t)$ given in Problem 7 and 8 in \S24 of Ref.~\cite{Landau1993}, we obtain to leading order
 \beq{avrPdot}
 \frac{1}{F_\mathrm{rad}}\:\Big\langle\frac{\dm P}{\dm t}\Big\rangle =
 \left\{ \begin{array}{rl}
 \dpst \frac{4}{2\rhoOTi+1} \frac{\dels}{a} \ll 1 , & \text{ for }\: a\gg\dels,
 \\[4mm]
 \dpst \frac{2}{\pi} \frac{a}{\dels} \ll 1 ,        &  \text{ for }\: a\ll\dels.
 \end{array}  \right.
 \eeq
We conclude that $\avr{\frac{\dm P}{\dm t}} \ll F_\mathrm{rad}$ in both the large and the small particle limit, and hence the assumption of \eqref{FradExactFF} is fulfilled in those limits.

Considering typical microparticle acoustophoresis experiments, the unsteady acceleration takes place on a timescale between micro- and milli-seconds, much shorter than the time of a full trajectory. Typically, the unsteady part of the trajectory is not resolved and it is not important to the experimentally observed quasi-steady particle trajectory. In acoustic levitation~\cite{Brandt2001,Xie2001,Vandaele2005,Foresti2014}, where there is no drift, the assumption of time periodicity is exact. We conclude that the assumption of time periodicity is not restricting practical applications of our theory.

\subsection{Perturbation expansion and linearity}

The second fundamental assumption of our theory  is the validity of the perturbation expansion, which requires the acoustic perturbation parameter $\veac$ of \eqref{veac_def} to be much smaller than unity. For applications in particle-handling in acoustophoretic microchips \cite{Barnkob2010, Augustsson2011}, this constraint is not very restrictive as typical resonant acoustic energy densities of 100~J/m$^{3}$ result in $\veac\sim10^{-4}$.

Given the validity of the linear first-order equations, the solutions we have obtained for $\fO$ and $\fI$ based on the particular incident plane wave $\phi_i = \phi_0\:\ee^{\ii \kc z}$ are general, since any incident wave at frequency $\omega$ can be written as a superposition of plane waves.

\begin{table*}[!t]
\caption{\tablab{parameter_values} Material parameter values at ambient pressure 0.1~MPa and temperature 300\:K used in this study, given for water (wa) \cite{Muller2014, Wagner2002, Huber2009, Huber2012}, an average liquid food oil \cite{Coupland1997}, air \cite{crcChemPhys}, and polystyrene (ps) \cite{OndaCorp, crcPolymers, Domalski1996, Chang1968}. Parameter values for water and oil at other temperatures are obtained from the fits in Refs.~\cite{Muller2014, Coupland1997}.}
\begin{ruledtabular}
\begin{tabular}{l c r@{$\:\times\:$}l  r@{$\:\times\:$}l  r@{$\:\times\:$}l  r@{$\:\times\:$}l c}
Parameter & Symbol & \multicolumn{2}{c}{Value (wa)} & \multicolumn{2}{c}{Value (oil)} & \multicolumn{2}{c}{Value (air)} & \multicolumn{2}{c}{Value (ps)} & Unit \\ \hline \\[-2.0mm]
Longitudinal speed of sound \hspace{2.5mm}
 &
$c$
 &
$1.502$ & $10^{3}$
 &
$1.445$ & $10^{3}$
 &
$3.474$ & $10^{2}$
 &
$2.40$ & $10^{3}$
 &
m\:s$^{-1}$
\\[1.0mm]
Transverse speed of sound
 &
$c^{{}}_\mathrm{T}$
 &
\multicolumn{2}{c}{$-$}
 &
\multicolumn{2}{c}{$-$}
 &
\multicolumn{2}{c}{$-$}
 &
$1.15$ & $10^{3}$
 &
m\:s$^{-1}$
\\[1.0mm]
Mass density
 &
$\rho_0$
 &
$9.966$ & $10^{2}$
 &
$9.226$ & $10^{2}$
 &
$1.161$ & $10^{0}$
 &
$1.05$ & $10^{3}$
 &
kg\:m$^{-3}$
\\[1.0mm]
Compressibility
 &
$\kappa^{{}}_s$
 &
\hspace{1.5mm} $4.451$ & $10^{-10}$ \hspace{1.5mm}
 &
\hspace{1.5mm} $5.192$ & $10^{-10}$ \hspace{1.5mm}
 &
\hspace{1.5mm} $7.137$ & $10^{-6}$ \hspace{1.5mm}
 &
\hspace{1.5mm} $2.38$ & $10^{-10}$ \hspace{1.5mm}
 &
Pa$^{-1}$
\\[1.0mm]
Thermal expansion coefficient \hspace{2.5mm}
 &
$\alpha^{{}}_p$
 &
$2.748$ & $10^{-4}$
 &
$7.046$ & $10^{-4}$
 &
$3.345$ & $10^{-3}$
 &
$2.09$ & $10^{-4}$
 &
K$^{-1}$
\\[1.0mm]
Specific heat capacity
 &
$c^{{}}_p$
 &
$4.181$ & $10^{3}$
 &
$2.058$ & $10^{3}$
 &
$1.007$ & $10^{3}$
 &
$1.22$ & $10^{3}$
 &
J\:kg$^{-1}$\:K$^{-1}$
\\[1.0mm]
Heat capacity ratio
 &
$\gamma$
 &
$1.012$ & $10^{0}$
 &
$1.151$ & $10^{0}$
 &
$1.402$ & $10^{0}$
 &
$1.04$ & $10^{0}$
 &
$1$
\\[1.0mm]
Shear viscosity
 &
$\eta_0$
 &
$8.538$ & $10^{-4}$
 &
$4.153$ & $10^{-2}$
 &
$1.854$ & $10^{-5}$
 &
\multicolumn{2}{c}{$-$}
 &
Pa\:s
\\[1.0mm]
Bulk viscosity\:\footnote{The bulk viscosity is negligible for scattering in the long-wavelength limit but has been included for completeness. Values for water, oil and air are estimated from Refs.~\cite{Holmes2011},~\cite{Chanamai1998}, and \cite{Prangsma1973}, respectively. For oil, $\eta_0^\mathrm{b}$ is obtained from the attenuation constant $\alpha_0$ at 298.15 K and 10 MHz \cite{Chanamai1998} using $\alpha_0 = 2\pi^2f^2/(\rho_0 c^3) [\eta_0^\mathrm{b} + (4/3)\eta_0 + (\gamma-1)\kth/c_p]$.}
 &
$\eta_0^\mathrm{b}$
 &
$2.4$ & $10^{-3}$
 &
$8.3$ & $10^{-2}$
 &
$1.1$ & $10^{-5}$
 &
\multicolumn{2}{c}{$-$}
 &
Pa\:s
\\[1.0mm]
Thermal conductivity
 &
$k_\mathrm{th}$
 &
$6.095$ & $10^{-1}$
 &
$1.660$ & $10^{-1}$
 &
$2.638$ & $10^{-2}$
 &
$1.54$ & $10^{-1}$
 &
W\:m$^{-1}$\:K$^{-1}$
\end{tabular}
\end{ruledtabular}
\end{table*}

\subsection{Oscillations of the suspended particle}
\seclab{OscParticle}

The third fundamental assumption of our theory is the assumption of small particle oscillation amplitudes, allowing the boundary conditions to be evaluated at the fixed interface position $r=a$. This assumption puts physical constraints on the volume oscillations, \figref{rad_force_concept}(a) and (b), and the center-of-mass oscillations, \figref{rad_force_concept}(c).

The volume oscillations of the particle are due to mechanical and thermal expansion. From the definition of the compressibility $\kappa'_s$ and the volumetric thermal expansion coefficient $\alpha'_p$, we estimate the maximum relative change in particle radius $\Delta a/a$ to be
 \bsub
 \eqlab{constraint_volume}
 \bal
 \frac{\Delta a}{a} &\simeq \frac{\kappa'_s}{3} p_1 = \frac{\tilde{\kappa}_s}{3} \veac \ll 1, \\
 \frac{\Delta a}{a} &\simeq \frac{\alpha'_p}{3} T_1 \simeq \frac{1}{3}\:(\gamma - 1) \alfpTi \veac \ll 1 .
 \eal
 \esub
Here, we have used $\kappa_s p_1 = \veac$ and $T_1= \frac{(\gamma-1)\kappa_s}{\alpha_p} p_1$ obtained from \eqref{thermodynamic_identities} in the adiabatic limit $s_1 = 0$ combined with \eqref{gamma_small}. Except for gas bubbles in liquids, for which $\kapsTi \gg 1$, these inequalities are always fulfilled for small perturbation parameters $\veac$.

The velocity of the center-of-mass oscillations is found from Eq. (37) of Ref.~\cite{Settnes2012} to be $v_\mathrm{p}^\mathrm{osc}= \frac{3}{2} \frac{f_1}{\rhoOTi-1}v_\mathrm{in}$. In the large-particle limit, $\fI$ is given by \eqref{f1_large}, which implies $0 < v_\mathrm{p}^\mathrm{osc} < 3 v_\mathrm{in}$, where the lower and the upper limit is for $\rhoOTi \gg 1$ and $\rhoTi \ll 1$, respectively. In the point-particle limit, \eqref{f1_point}, $v_\mathrm{p}^\mathrm{osc}=v_\mathrm{in}$ independent of $\rhoOTi$. The relative displacement amplitude $\Delta \ell / a$ is hence estimated as
 \beq{constraint_displacement}
 \frac{\Delta\ell}{a} \simeq \dfrac{v_\mathrm{p}^\mathrm{osc}}{\omega a} \simeq
 \left\{\!\!
 \begin{array}{rll}
 \dpst \frac{3}{2\rhoOTi+1} \frac{\lambda}{2\pi a} \veac \!\!&\!\!\ll 1, & \text{for }\; a \gg \dels, \\[4mm]
 \dpst \frac{\lambda}{2\pi a} \veac \!\!&\!\!\ll 1, & \text{for }\; a \ll \dels,
 \end{array}
 \right.
 \eeq
and thus the general requirement is that $\veac \ll 2\pi a / \lambda$. For large particles in typical experiments, this restriction is not severe. However, for small particles it can be restrictive. For example, to obtain $\Delta \ell / a < 0.05$, we find for particles of radius $a=100~\SInm$ in water at 1~MHz and particles of radius $a=1~\SImum$ in air at 1~kHz, that $\veac \lesssim 10^{-5}$ and $\veac \lesssim 10^{-6}$, respectively.

\section{Microparticles and droplets in standing plane waves}
\seclab{examples}

The special case of a one-dimensional (1D) standing plane wave is widely used in practical applications of the acoustic radiation force in microchannel resonators \cite{Bruus2011, Barnkob2010, Thevoz2010, Augustsson2011, Grenvall2009, Liu2012, Hammarstrom2012, Scmid2014, Antfolk2014, Carugo2014, Shields2014, Leibacher2015, Peng2015} and acoustic levitators \cite{Brandt2001, Xie2001, Vandaele2005, Foresti2014}. The many application examples as well as its relative simplicity, makes the 1D case an obvious and useful testing ground of  our theory. In the following, we illustrate the main differences between our full thermoviscous treatment and the more conventional ideal-fluid or viscous-fluid models using the typical parameter values listed in \tabref{parameter_values}.

We consider a standing plane wave of the form $p_\mathrm{in}=p_\mathrm{a} \cos(k_0 y)$, $\vec{v}_\mathrm{in}=
\frac{\ii}{\rho_0 c}\: p_\mathrm{a} \sin(k_0 y) \vec{\mathrm{e}}_y$, with acoustic energy density $\Eac=\frac14\kappa_s p_\mathrm{a}^2 = \frac14\rho_0 v_\mathrm{a}^2$, where $p_\mathrm{a}$ and $v_\mathrm{a}$ are the pressure and the velocity amplitude, respectively. Expression~\eqrefnoEq{Frad_Settnes} for the radiation force then simplifies to
\bsub
\eqlab{Phiac}
\bal
\vec{F}_\mathrm{1D}^\mathrm{rad} &= 4 \pi \: \Phiac a^3 k_0 \Eac \sin(2 k_0 y) \vec{\mathrm{e}}_y , \\
\Phiac &= \dfrac{1}{3} \mathrm{Re} \left[f_0\right] + \dfrac{1}{2} \mathrm{Re}\left[f_1\right] ,
\eal
\esub
where $\Phiac$ is the so-called acoustic contrast factor. The radiation force is thus proportional to $\Phiac$, which contains the effects of thermoviscous scattering in $\fO$ and $\fI$. Note that for positive acoustic contrast factors, $\Phiac>0$, the force is directed towards the pressure nodes of the standing wave, while for negative acoustic contrast factors, $\Phiac<0$, it is directed towards the anti-nodes.

The acoustic contrast factor $\Phiac$ may be evaluated directly for an arbitrary particle size by using the expressions for the scattering coefficients, either $\fOfl$ and $\fIfl$ for a fluid droplet or $\fOsl$ and $\fIsl$ for a solid particle. For ease of comparison to the work of King~\cite{King1934}, Yosioka and Kawasima~\cite{Yosioka1955}, and Doinikov~\cite{Doinikov1997a, Doinikov1997b, Doinikov1997c}, we give the expression for the acoustic contrast factor $\Phiac^\mathrm{fl}$ of a fluid droplet for small boundary layers and in the point-particle limit. In the small-width boundary layer limit one obtains
 \bsub
 \eqlab{Phiac_fl_small_bnd}
 \bal
 \Phiac^\mathrm{fl} &= \frac{1}{3} \left( \frac{5\rhoOTi - 2}{2\rhoOTi+1}-\tilde{\kappa}_s \right) +
 \frac{3}{1+\tilde{\nu}_0^{1/2}\etaOTi^{\:-1}} \left( \frac{\rhoOTi - 1}{2\rhoOTi + 1} \right)^2 \frac{\dels}{a} \nn \\
 &\quad - \frac{1}{2} \frac{\gamma - 1}{1 + \DthTi^{1/2}\kthTi^{-1}} \left(1 - \frac{\tilde{\alpha}_p}{\rhoOTi \cpTi}\right)^2 \frac{\delt}{a} ,
 \\ &\nn \hspace*{8em} \text{(Small-width boundary layers)},
 \\
 &|\xc|^2 , |\xc^{\prime}|^2 \ll 1 \ll
 |\xs|^2 , |\xs^\prime|^2 , |\xt|^2 , |\xt^{\prime}|^2.
 \eal
 \esub
The first term is the well-known result given by Yosioka and Kawasima~\cite{Yosioka1955}, which reduces to that of King~\cite{King1934} for incompressible particles for which $\tilde{\kappa}_s=0$. The second term is the viscous correction, which agrees with the result of Settnes and Bruus~\cite{Settnes2012} for infinite particle viscosities, but extends it to finite particle viscosities. Note that the viscous correction yields a positive contribution to the acoustic contrast factor, while the thermal correction from the third term is negative. The result given in \eqref{Phiac_fl_small_bnd} is in agreement with the expression for the radiation force in a standing plane wave given by Doinikov~\cite{Doinikov1997c} in the weak dissipative limit of small boundary layers. However, this is not seen without considerable effort combining and reducing a number of equations. Although we find Doinikov's approach rigorous, it lacks transparency and is difficult to apply with confidence.

\begin{figure*}[!!t]
\centering
\includegraphics[width=1.0\textwidth]{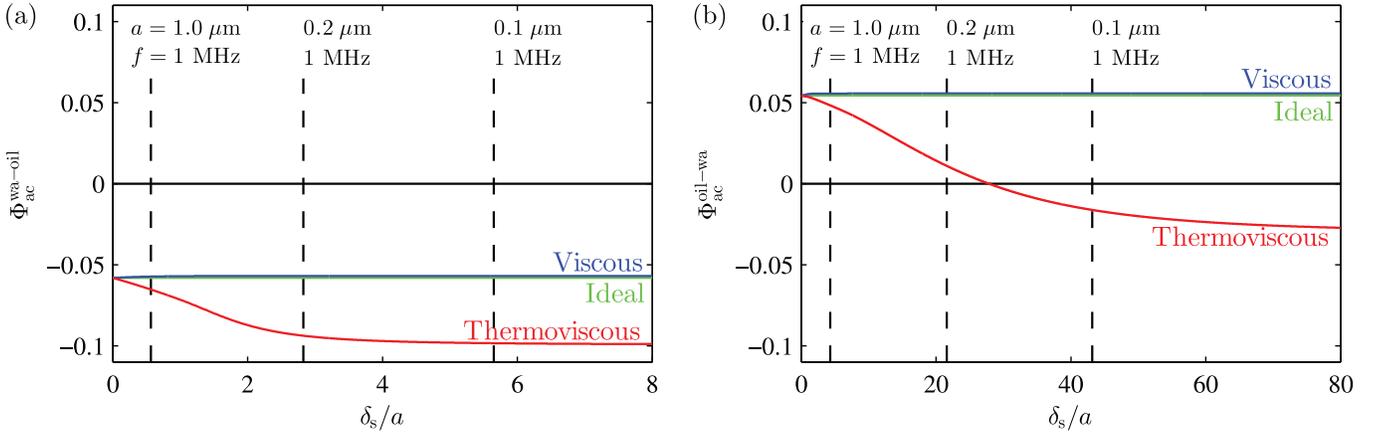}
\caption[]{(Color online) Acoustic contrast factor $\Phiac$ plotted as a function of $\dels/a$, the viscous boundary layer thickness in the medium normalized by particle radius. The curves are calculated using ideal theory (green), viscous theory (blue), and thermoviscous theory (red) for (a) an oil droplet in water (wa-oil) and (b) a water droplet in oil (oil-wa), both at 20$^\circ\mathrm{C}$. Thermoviscous theory leads to corrections to the acoustic radiation force around $100\%$. The vertical dashed lines indicate examples of particle sizes corresponding to the given value of $\dels/a$ at $f=1\: \mathrm{MHz}$. Note that the acoustic contrast factor changes sign at a critical particle radius for the case of water droplets in oil considered in (b).}
\figlab{Phi_wa-oil}
\end{figure*}

In the point-particle limit of infinitely large boundary layer thicknesses compared to the particle size, we obtain
 \bsub
 \eqlab{Phiac_fl_point}
 \bal
 &\Phiac^\mathrm{fl} = \frac{1}{3} \left[ (1\!-\!\tilde{\kappa}_s) - (1\!-\!\rhoOTi)
  - (\gamma\!- \!1) \rhoOTi \cpTi \Big(1 \!- \!\frac{\tilde{\alpha}_p}{\rhoOTi \cpTi}\Big)^2 \right] ,
 \\ &\nn \hspace*{14em} \text{(Point-particle limit)},
 \\
 &|\xc|^2, |\xc^{\prime}|^2 \ll 1 \text{ and }
 |\xs|^2 , |\xs^\prime|^2 , |\xt|^2 , |\xt^{\prime}|^2 \rightarrow 0,
 \eal
 \esub
in agreement with the viscous result of Settnes and Bruus~\cite{Settnes2012}, when omitting the last term stemming from thermal effects. The result for $\Phiac^\mathrm{fl}$ in~\eqref{Phiac_fl_point} is written in a form which emphasizes how parameter contrasts between particle and fluid lead to scattering. As expected, for $\tilde{\kappa}_s=1$ and $\rhoOTi=1$, the scattering due to compressibility and density (inertia) mechanisms vanishes. This is true for large particles~\cite{King1934,Yosioka1955,Gorkov1962,Settnes2012}, and it is reasonable that it remains true in the point-particle limit. The expressions for the acoustic radiation force on a point-particle in a standing plane wave given by Doinikov~\cite{Doinikov1997a,Doinikov1997b,Doinikov1997c} do not have this property, which is likely due to a sign-error or a misprint in the term corresponding to our dipole scattering coefficient $\fI$ in the point-particle limit~\eqref{f1_point}, as was also suggested by Settnes and Bruus~\cite{Settnes2012}.

The small-width boundary layer limit and the point-particle limit are useful for analyzing consequences of thermoviscous scattering on the acoustic radiation force, but we emphasize that our theory is not restricted to these limits. In general, the scattering coefficients $\fO$ and $\fI$ are functions of the non-dimensionalized wavenumbers $\xs$, $\xt$, $\xsp$, and $\xtp$. These may all be expressed in terms of the particle radius $a$ normalized by the thickness of the viscous boundary layer in the medium $\dels$,
\bsub
\begin{alignat}{2}
\xs  &= (1+\ii)\: \dfrac{a}{\dels} , &
\xsp &= (1+\ii) \sqrt{\dfrac{\rhoOTi}{\etaOTi}}\: \dfrac{a}{\dels} , \\
\xt  &= (1+\ii) \sqrt{\mathrm{Pr}}\: \dfrac{a}{\dels} , & \quad
\xtp &= \dfrac{(1+\ii)}{\sqrt{1-X'}} \sqrt{\dfrac{\mathrm{Pr}}{\tilde{D}_\mathrm{th}}}\: \dfrac{a}{\dels} ,
\end{alignat}
\esub
where we have used $\dels' = \dels \sqrt{\etaOTi/\rhoOTi}$, $\delt =  \dels \sqrt{1/\mathrm{Pr}}$,
$\delt' =  \dels \sqrt{\big[(1-X')\DthTi\big]/\mathrm{Pr}}$, with $\mathrm{Pr}=\nu_0 / \Dth$ being the Prandtl number of the fluid medium and $X'$ set to zero for the fluid droplet case. Below, we investigate the thermoviscous effects on the acoustic radiation force by plotting the acoustic contrast factor $\Phiac$ as a function of $\dels/a$, ranging from zero boundary-layer effects at $\dels /a=0$ to maximum effects in the limit $\dels/a \rightarrow \infty$.

\begin{figure*}[!!t]
\centering
\includegraphics[width=1.0\textwidth]{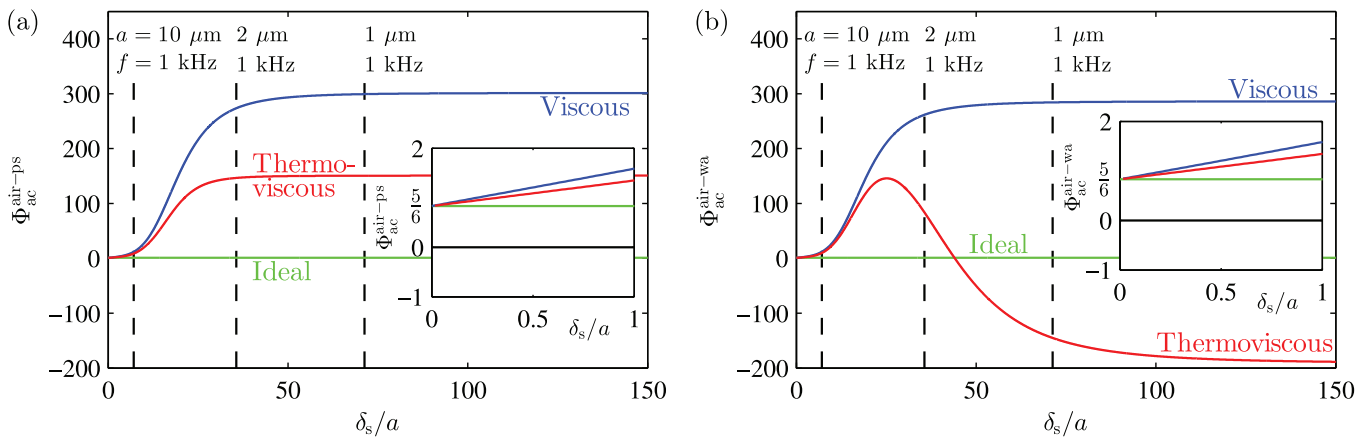}
\caption[]{(Color online) Acoustic contrast factor $\Phiac$ for particles in air plotted as a function of $\dels/a$, the viscous boundary layer thickness in the medium normalized by particle radius. The curves are calculated using ideal theory (green), viscous theory (blue), and thermoviscous theory (red) for (a) a polystyrene particle in air (air-ps) and (b) a water droplet in air (air-wa), both at 300 K. Ideal theory predicts a constant value of $\Phiac = 5/6$ independent of particle size.  For particles much smaller than the boundary layer thickness, however, thermoviscous theory predicts huge deviations from ideal theory leading to acoustic contrast factors two orders of magnitude larger than expected from ideal-fluid theory. The vertical dashed lines indicate examples of particle sizes corresponding to the given value of $\dels/a$ at $f=1~\SIkHz$.}
\figlab{Phi_air}
\end{figure*}

\subsection{Oil droplets in water and water droplets in oil}

We first consider the cases of water with a suspended oil droplet (wa-oil) and of oil with a suspended water droplet (oil-wa) using the parameters of a typical food oil given in \tabref{parameter_values}. Since the density contrast of water and oil is small, the dipole scattering with its viscous effects is small, while on the other hand the thermal effects in the monopole scattering are significant. This is clearly seen from \figref{Phi_wa-oil}, where the acoustic contrast factor $\Phiac$ is plotted for the two cases as function of $\dels/a$ using ideal theory, viscous theory, and full thermoviscous theory. \figref{Phi_wa-oil} shows that for sub-micrometer droplets at MHz frequency the thermoviscous theory leads to corrections around 100$\%$ as compared to the ideal and the viscous theory, which manifestly demonstrates the importance of thermal effects in such systems.

We note from \figref{Phi_wa-oil}(a) that the acoustic contrast factor of oil droplets in water is negative, which means that oil droplets are focused at the pressure anti-nodes. Conversely, water droplets in oil are thus expected to be focused at the pressure nodes. However, in \figref{Phi_wa-oil}(b) we see that thermoviscous theory predicts a tunable sign-change in the acoustic contrast factor  as a result of the negative thermal corrections to the monopole scattering coefficient. This means that droplets above a critical size threshold experience a force directed towards the pressure nodes, while droplets smaller than the threshold experience a force towards the anti-nodes, even though the only distinction between the droplets is their size. This sign-change in $\Phiac$ can also be achieved for elastic solid particles under properly tuned conditions. By changing, for example, the compressibility contrast $\tilde{\kappa}_s$, the curves for $\Phiac(\dels/a)$ may be shifted vertically and a possible size-threshold condition may be changed. Moreover, since $\dels =\sqrt{2\eta_0/(\rho_0 \omega)}$ and $\delt=\sqrt{2\kth/(\rho_0 c_p \omega)}$, there are several direct ways of tuning a threshold value, e.g. by frequency or by changing the density of the medium.

\subsection{Polystyrene particles and water droplets in air}
\seclab{ps_wa_in_air}

Using the particular cases of a polystyrene particle and a water droplet suspended in air as main examples, we study the effects of a large density contrast $\rhoOTi \gg 1$, for which our thermoviscous theory predicts much larger radiation forces on small particles than ideal-fluid theory, for which $\Phiac^\mathrm{ideal} = 5/6$ independent of particle size. This is demonstrated in \figref{Phi_air}, where $\Phiac$ is plotted as a function of $\dels/a$ for the two particle types. In the large-particle limit $\dels/a = 0$, boundary-layer effects are negligible, and ideal, viscous, and thermoviscous theory predict the same contrast factor $\Phiac = 5/6$, but as $\dels/a$ increases, the thermoviscous and viscous theory predict an increased value of $\Phiac$, approximately $2\Phiac^\mathrm{ideal}$ for $\dels/a = 1$ as seen in the insets of \figref{Phi_air}(a) and (b). Decreasing the particle size further, $\dels/a \gg 1$, the thermoviscous effects become more pronounced with $\Phiac/\Phiac^\mathrm{ideal} \sim 10^2$. Choosing the frequency to be 1~kHz, this remarkable deviation from ideal-fluid theory is obtained for moderately-sized particles of radius $a \approx 2~\SImum$.

While $\Phiac^\mathrm{air-ps}$ in \figref{Phi_air}(a) for the polystyrene particle is a monotonically increasing function of $\dels/a$, the $\Phiac^\mathrm{air-wa}$ in \figref{Phi_air}(b) of a water droplet exhibits a non-monotonic behavior. For small values of $\dels/a \lesssim 25$, the viscous dipole scattering dominates resulting in a positive contrast factor $\Phiac^\mathrm{air-wa} \lesssim 10^2$. For larger values, $\dels/a \gtrsim 25$, thermal effects in the monopole scattering become dominant leading to a sign-change in $\Phiac^\mathrm{air-wa}$ and finally to large negative contrast factors  approximately equal to $-10^2$ as the point-particle limit $\dels/a \gg 1$ is approached. This example clearly demonstrates how the acoustic contrast factor may exhibit a non-trivial size-dependency with profound consequences for the acoustic radiation force on small particles. The detailed behavior depends on the specific materials but can be calculated using \eqref{Phiac} and the expressions for $\fO$ and $\fI$ listed in \tabref{eqs_overview}.

\section{Conclusion}

Since the nominal work of Epstein and Carhart~\cite{Epstein1953} and Allegra and Hawley~\cite{Allegra1972}, effects of thermoviscous scattering have been known to be important for ultrasound attenuation in emulsions and suspensions of small particles. In this paper, we have by theoretical analysis shown that thermoviscous effects are equally important for the acoustic radiation force $\FFFrad$  on a small particle. $\FFFrad$ is evaluated from \eqref{Frad_Settnes}, or more generally from \eqref{Frad_Settnes_freq}, using our new analytical results for the thermoviscous scattering coefficients $\fO$ and $\fI$  summarized in \tabref{eqs_overview}. Our analysis places no restrictions on the viscous and thermal boundary layer thicknesses $\dels$ and $\delt$ relative to the particle radius $a$, a point which is essential to calculation of the acoustic radiation force on micro- and nanometer-sized particles.

The discussion in \secref{Frad1} leading to \eqref{Frad_Settnes} for $\FFFrad$, as well as the discussion of the range of validity presented in \secref{constraints}, are intended to provide clarification and a deeper insight into the fundamental assumptions of the theory for the acoustic radiation force. Foremost, we have extended the discussions of the role of streaming, the fundamental assumption of time periodicity, and the trick of evaluating the radiation force in the far-field. To our knowledge, the exact non-perturbative expression~\eqrefnoEq{FradExactFF} for the radiation force evaluated in the far-field has not previously been given in the literature.

For the simple case of a 1D standing plane wave at a single frequency, the expression~\eqrefnoEq{Frad_Settnes_freq} for $\FFFrad$ simplifies to the useful expression given in \eqref{Phiac}, which involves the acoustic contrast factor $\Phiac$. Similar simplified expressions can be derived for other cases of interest such as that of a 1D traveling plane wave. An important result from the discussion of the simple 1D case in \secref{examples} is that we must abandon the notion of a purely material-dependent acoustic contrast factor $\Phiac$. In general, $\Phiac$ also depends on the particle size, and in many cases this size-dependency can even lead to a sign change in $\Phiac$ at a critical threshold. Recent acoustophoretic experiments on sub-micrometer-sized water droplets and smoke particles in air may provide the first evidence of this prediction~\cite{Ran2015}. Considering only viscous corrections, however, the authors could not fully explain their data. Our analysis suggests that thermoviscous effects must be taken into account when designing and analyzing such experiments.

Our results for the acoustic radiation force in a standing plane wave evaluated using \eqref{Phiac} agree with the expressions obtained from the work of Doinikov~\cite{Doinikov1997a,Doinikov1997b,Doinikov1997c} in the limit of small boundary layers, but not in the opposite limit of a point particle. In our theory both of these limits are evaluated directly using the derived analytical expressions valid for arbitrary boundary layer thicknesses, and we have furthermore given a physical argument supporting our result in the point-particle limit. Considering the viscous theory of Danilov and Mironov~\cite{Danilov2000}, we remark that their result is based on the viscous reaction force on an oscillating rigid sphere~\cite{Landau1993} instead of a direct solution of the governing equations for an acoustic field scattering on a sphere.

Importantly, we have shown that the acoustic radiation force on a small particle including thermoviscous effects may deviate by orders of magnitude from the predictions of ideal-fluid theory when there is a large density contrast between the particle and the fluid. This result is particularly relevant for acoustic levitation and manipulation of small particles in gases~\cite{Brandt2001, Xie2001, Vandaele2005, Foresti2014}. Thermoviscous effects can also be significant in many lab-on-a-chip applications involving ultrasound handling of submicrometer-sized particles such as bacteria and vira~\cite{Hammarstrom2012, Antfolk2014}.

A firm theoretical understanding of thermoviscous effects, and of the particle-size-dependent sign change of the acoustic contrast factor, could prove important for future applications relying on ultrasound manipulation of micro- and nanometer-sized particles.

%%%%%%%%%%%%%%%%%%%%%%%%%%%%%%% Appendix %%%%%%%%%%%%%%%%%%%%%%%%%%%%%%%%%%

\appendix

\section{Velocity and normal stress in spherical coordinates}
\seclab{stress_tensor_app}

In spherical coordinates $(r,\theta,\varphi)$ with azimuthal symmetry, using that $\vvvI = \nablabf \phi + \rot\vpsi$ with  $\phi = \phic + \phit$ and $\vpsi = \psis \: \textbf{e}_{\varphi}$, the first-order velocity components are
\bsub
\eqlab{velocity_spherical_coords}
\bal
v_{1r} &= \pp_r \phi + \dfrac{1}{r \sin\theta} \pp_{\theta}\big[ \sin\theta \: \psis \big] , \\
v_{1\theta} &= \dfrac{1}{r} \pp_{\theta} \phi - \dfrac{1}{r} \pp_r \big[ r \: \psis \big].
\eal
\esub
Inserting this into \eqref{stress_field_fl}, we obtain the normal components of the first-order stress tensor
 \bsub
 \eqlab{stress_tensor_spherical_sl}
 \bal
 \sigma_{1rr} &= \eta_0 (2 \kc^2 - \ks^2) \phic + \etaO (2\kt^2 - \ks^2) \phit
 + 2\etaO \ppsqr_r \phi
 \nn \\
 & \qquad
 +  \dfrac{2\etaO}{\sin\theta} \pp_{\theta} \bigg[ \sin\theta \left( \dfrac{1}{r} \pp_r \psis - \dfrac{1}{r^2} \psis \right) \bigg], \\[2mm]
 \sigma_{1 \theta r} &= 2\eta_0 \pp_{\theta} \bigg(\dfrac{1}{r} \pp_r \phi
 - \dfrac{1}{r^2}\phi \bigg) - \etaO \left( \ppsqr_r \psis - \dfrac{2}{r^2} \psis \right)
 \nn \\
  & \qquad
  + \dfrac{\etaO}{r^2} \pp_{\theta} \bigg[ \dfrac{1}{\sin\theta} \pp_{\theta}
  \big(\sin\theta \: \psis \big) \bigg].
 \eal
 \esub

\section{The scattering coefficients $\bm{\fO}$ and $\bm{\fI}$}
\seclab{app_scattering_coeffs}
Here, we outline the calculation of the monopole and dipole scattering coefficients $f_0$ and $f_1$ in the long-wavelength limit where the particle radius and the boundary layer thicknesses are assumed much smaller than the wavelength. Defining the small parameter $\ve=k_0 a\ll 1$, we note that $k_0 a, k_0 \dels, k_0 \delt, k_0 \delt^\prime$, and for a fluid particle furthermore $k_0 \dels^\prime$, are all of order $\ve$. The calculation is carried out to first order in $\ve$.

\subsection{The monopole scattering coefficient $\bm{\fO}$}
\seclab{monopole_calculations}
The monopole scattering coefficient $f_0$ may be obtained from Eqs.~\eqrefnoEq{bc_eqs_a}, \eqrefnoEq{bc_eqs_c}, \eqrefnoEq{bc_eqs_d} and \eqrefnoEq{bc_eqs_f} setting $n=0$ and $C_0=C_0^\prime=0$. All Bessel functions of the small arguments $\xc, \xcp \sim \ve \ll 1$ are expanded to first order in $\ve$ using \eqref{bessel_approx} of  \appref{special_funcs}, and in the (unprimed) fluid medium we neglect $\xc^2$ in comparison to $\xs^2$. Thus, we arrive at
 \bsub
 \eqlab{monopole_bc_eqs_B}
 \bal
 & A_0 \dfrac{\ii}{\xc} + A_0^{\prime} \dfrac{1}{3} \xc^{\prime 2} - B_0 \xt h_1(\xt)
 + B_0^{\prime} \xtp j_1(\xtp) = \dfrac{1}{3} \xc^2 , \eqlab{monopole_bc_eqs_Ba}
 \\
 & A_0 \bc \left( 1 - \dfrac{\ii}{\xc}\right) - A_0^{\prime} \bcp + B_0 \bt h_0(\xt)  - B_0^{\prime} \btp j_0(\xtp) = - \bc ,
 \\
 & A_0 \kth \bc \dfrac{\ii}{\xc} + A_0^{\prime} \dfrac{1}{3} \kth^{\prime} \bcp \xc^{\prime 2} - B_0 \kth \bt \xt h_1(\xt) \nn \\
 &\quad\quad + B_0^{\prime} \kth^{\prime} \btp \xtp j_1(\xtp) = \dfrac{1}{3} \kth \bc \xc^2 , \eqlab{monopole_bc_eqs_Bc}
 \eal
 \bal
 & A_0 \eta_0 \left[ (4-\xs^2) \dfrac{\ii}{\xc} + \xs^2 \right] - A_0^{\prime} \eta_0^{\prime} \left[ \xs^{\prime 2} - \dfrac{4}{3} \xc^{\prime 2} \right] \nn \\
 &\quad\quad + B_0 \eta_0 \left[ ( \xs^2 - 2 \xt^2 ) h_0(\xt) - 2 \xt^2 h_0^{\prime\prime}(\xt) \right] \nn \\
 & \quad\quad - B_0^{\prime} \eta_0^{\prime} \left[ (\xs^{\prime 2} - 2 \xt^{\prime 2} ) j_0(\xtp) - 2 \xt^{\prime 2} j_0^{\prime\prime}(\xtp) \right] = - \eta_0 \xs^2 ,
 \eal
 \esub
where \eqref{bessel_rec} is used to write $g_0^{\prime}(x)=-g_1(x)$ for any spherical Bessel of Hankel function  $g_0(x)$.

Multiplying \eqref{monopole_bc_eqs_Bc} by $1/(\kth \bt)$ and using the ratios
 \begin{alignat}{2}
  \eqlab{b_ratios}
 \frac{b_c}{b_t} &= -(\gamma-1)\frac{\xc^2}{\xt^2},
 \qquad &
 \frac{b'_c}{b_c} &= \chiTi\frac{\alfpTi}{\tilde{c}_p},
 \\ \nn
 \frac{b'_t}{b_t} &=  \frac{1}{\chiTi\alfpTi \DthTi},
 \qquad &
 \frac{b'_c}{b_t} &=  \frac{b_c}{b_t} \frac{b'_c}{b_c} =
 -\chiTi(\gamma-1)\frac{\alfpTi}{\tilde{c}_p}\frac{\xc^2}{\xt^2},
 \end{alignat}
of the $b$-coefficients defined in \eqref{T1_bc_bt} (here, $\chiTi = 1$ for a droplet and $\chiTi = \chi'$ for a solid particle, respectively,  while Eqs.~\eqrefnoEq{gamma_small}, \eqrefnoEq{c_rho0_kaps}, and \eqrefnoEq{chi_def} is used to reduce $b'_c/b_c$), we note that the $A_0$ and $A'_0$ terms can be neglected to order $\ve$, and we obtain
 \beq{B0_B0p}
 B_0^{\prime} = \frac{\kth \bt}{\kth^{\prime} \btp}\: \frac{\xt h_1(\xt)}{\xtp j_1(\xtp)}\: B_0
 = \chiTi\:\frac{\alfpTi}{\rhoOTi\tilde{c}_p}\: \frac{\xt h_1(\xt)}{\xtp j_1(\xtp)}\: B_0.
 \eeq
With this, we eliminate $B_0^{\prime}$ from the system of equations~\eqrefnoEq{monopole_bc_eqs_B}, and the remaining three equations become
 \beq{monopole_bc_eqs_C}
 \left(\!\!
 \begin{array}{ccc}
 \frac{\ii}{\xc} & \frac{1}{3} \xc^{\prime 2} & -S_1\\[2mm]
 \frac{\bc}{\bt}\big( \frac{\ii}{\xc}-1\big) & \frac{\bcp}{\bt} & - \frac{S_2}{\xt^2} \\[2mm]
 \frac{\ii(\xs^2-4)}{\xc} - \xs^2 &
 \big(\xspsqr-\frac{4}{3}\xc^{\prime 2}\big)\etaOTi& -S_3
 \end{array}
 \!\!\right)\!
 \left(\!\begin{array}{c}  A_0 \\[3mm] A'_0 \\[3mm] B_0  \end{array} \!\right)
 =
 \left(\! \begin{array}{c}
 \frac{\xc^2}{3} \\[3mm]
 \frac{\bc}{\bt}\\[3mm]
 \xs^2
 \end{array} \!\right)\!,
 \eeq
where we have introduced the functions $S_1$, $S_2$, and $S_3$,
 \bsub
 \eqlab{functions_S}
 \bal
 S_1 &= \Bigg[ 1 - \frac{1}{\kthTi}\frac{\bt}{\btp} \Bigg] \xt h_1(\xt) , \\
 S_2 &= \xt^2\Bigg[\frac{h_0(\xt)}{\xt h_1(\xt)} - \frac{1}{\kthTi} \dfrac{j_0(\xtp)}{\xtp j_1(\xtp)}\Bigg]
 \xt h_1(\xt) , \\ \nn
 S_3 &= \Bigg[\frac{\xs^2 h_0(\xt)}{\xt h_1(\xt)} - 4 \left( 1 - \frac{\etaOTi}{\kthTi} \dfrac{\bt}{\btp} \right)
 - \frac{\etaOTi}{\kthTi} \dfrac{\bt \xspsqr}{\btp \xtp} \dfrac{j_0(\xtp)}{j_1(\xtp)}\Bigg]
 \\
 & \quad
  \times \xt h_1(\xt),
 \eal
 \esub
and the relative shear constant $\etaOTi$ obtained from \eqref{ks_flsl_general},
 \beq{etaOTiDef}
 \etaOTi = \frac{\eta'_0}{\eta_0} = \rhoOTi\:\frac{\xs^2}{\xspsqr}.
 \eeq
In obtaining the expression for $S_3$ we have used \eqref{bessel_rec} to substitute $g_0^{\prime\prime}(x)=-g_0(x) + (2/x) g_1(x)$ for any spherical Bessel or Hankel function $g(x)$. Using Eq.~\eqrefnoEq{b_ratios}, \eqrefnoEq{etaOTiDef}, and the explicit forms~\eqrefnoEq{bessel_explicit} of the Bessel functions, the $S$-functions are expressed in terms of the dimensionless wavenumbers as
 \bsub
 \eqlab{functions_S_reduced}
 \bal
 S_1 &=
 \bigg[1-\chiTi \frac{\alfpTi}{\rhoOTi\cpTi}\bigg] \xt h_1(\xt),
 \\
 S_2 &=\frac{1}{H(\xt,\xtp)}\: \xt h_1(\xt),
 \\
 S_3 &=
 \bigg[\!\frac{\xs^2}{1-\ii\xt} - 4 +
 \frac{\chiTi \alfpTi}{\cpTi}\bigg(
 \frac{4\xs^2}{\xspsqr}
 \!-\!\frac{ \xs^2 \tan \xtp}{\tan\xtp - \xtp} \bigg)\!\bigg] \xt h_1(\xt),
 \eal
 \esub
where $H(\xt,\xtp)$ is given in \eqref{monopole_H}. The coefficient $A_0$ is now found from \eqref{monopole_bc_eqs_C} by the method of determinants (Cramer's rule) as $A_0  = D(A_0)/D$, where $D$ is the determinant of the left-hand-side system matrix and $D(A_0)$ is determinant of the system matrix in which the first column (the $A_0$ coefficients) are replaced by the right-hand-side column with the inhomogeneous terms. The monopole scattering coefficient $f_0$ in the long-wavelength limit can then be expressed as
\bal
\eqlab{f0_general_app}
f_0 =  \frac{3\ii}{\xc^3}\:A_0  = \frac{3\ii}{\xc^3} \dfrac{D(A_0)}{D},
\eal
with the determinants $D$ and $D(A_0)$ given by
 \bsub
 \eqlab{D_DA0}
 \bal
 D &= - S_1 \Bigg[\etaOTi\dfrac{\bc}{\bt} \left( \dfrac{\ii}{\xc} - 1 \right) \left( \dfrac{4}{3} \xc^{\prime 2} - \xs^{\prime 2} \right)
 \nn \\
 & \qquad \qquad - \dfrac{\bcp}{\bt} \left( (4- \xs^2) \dfrac{\ii}{\xc} + \xs^2 \right) \Bigg]
 \nn \\
 & \quad
 - \dfrac{S_2}{\xt^2} \left[\dfrac{\xc^{\prime 2}}{3}  \left(  \dfrac{\ii(4 - \xs^2)}{\xc} + \xs^2 \right) - \dfrac{\ii \etaOTi}{\xc} \left( \dfrac{4}{3} \xc^{\prime 2} - \xs^{\prime 2} \right) \right]
 \nn \\
 & \quad - S_3 \left[ \dfrac{1}{3} \xc^{\prime 2} \dfrac{\bc}{\bt} \left( \dfrac{\ii}{\xc} - 1 \right) - \dfrac{\ii}{\xc} \dfrac{\bcp}{\bt} \right],
 \eal
 \bal
 D(A_0) &= - S_1 \left[ \etaOTi\dfrac{\bc}{\bt} \left( \dfrac{4}{3} \xc^{\prime 2} - \xs^{\prime 2} \right) + \dfrac{\bcp}{\bt} \xs^2 \right]
 \nn \\
 & \quad
 - \dfrac{S_2}{3 \xt^2} \left[ - \etaOTi\xc^2 \left( \dfrac{4}{3} \xc^{\prime 2} - \xs^{\prime 2} \right) - \xc^{\prime 2} \xs^2 \right]
 \nn \\
 & \quad
 - \dfrac{S_3}{3} \left[ \dfrac{\bc}{\bt} \xc^{\prime 2} - \dfrac{\bcp}{\bt} \xc^2 \right] .
 \eal
 \esub
The solution $A_0 = D(A_0)/D$, though written somewhat differently, agrees with Allegra and Hawley's Eq.~(10) of Ref.~\cite{Allegra1972}.

\subsubsection{$f_0$ for a suspended thermoviscous droplet}
For a suspended thermoviscous droplet, the precise definition of the long-wavelength limit is given in \eqref{fluid-fluid_longwave}. In this case, the shear mode characterized by $\xsp$ inside the droplet corresponds to a boundary layer, and consequently comparison to the compressional mode inside and outside the droplet yields $\xc^2/\xspsqr \sim \xcpsqr/\xspsqr \sim \ve^2 \ll 1$. This, combined with $\bc/\bt \sim \bcp / \bt \sim \xc^2/\xt^2 \sim \ve^2 \ll 1$ from \eqref{b_ratios}, leads to the following simplification of \eqref{D_DA0} to first order in $\ve$,
 \bsub
 \bal
 D &\simeq - \dfrac{\ii}{\xc} \dfrac{\xs^2}{\xt^2} \rhoOTi S_2 ,
 \\
 D(A_0) &\simeq - \dfrac{\rhoOTi}{3} \dfrac{\xs^2}{\xt^2}
 \Bigg( \xc^2 - \dfrac{\xc^{\prime 2}}{\rhoOTi}\Bigg) S_2
 + \rhoOTi \xs^2 \dfrac{\bc}{\bt}
 \Bigg(  1 - \dfrac{\alfpTi}{\rhoOTi \tilde{c}_p} \Bigg) S_1.
 \eal
 \esub
When inserting this into \eqref{f0_general_app}, we obtain
\bal
\fOfl = 1 - \tilde{\kappa}_s + 3 (\gamma - 1) \left( 1 - \dfrac{\alfpTi}{\rhoOTi \tilde{c}_p} \right) \dfrac{S_1}{S_2},
\eal
which upon substitution with $\frac{S_1}{S_2} = \big(1 - \frac{\alfpTi}{\rhoOTi \cpTi} \big)H(\xt,\xtp)$  from \eqref{functions_S_reduced} with $\chiTi = 1$, leads to the final analytical result for $\fOfl$ given in \eqref{f0_fluid-fluid}.

\subsubsection{$f_0$ for a suspended thermoelastic particle}
The qualitative change going from the thermoviscous droplet to the thermoelastic particle lies in the shear mode, which changes from a highly damped boundary layer mode to a propagating transverse wave with $\xspsqr \sim \ve^2$. A further implication is that the shear constant ratio of \eqref{etaOTiDef} becomes large, $\etaOTi = \rhoOTi \xs^2/\xspsqr \sim \ve^{-2} \gg 1$, and order of magnitude wise, the $S$-functions of \eqref{functions_S_reduced} obey $S_1 \sim S_2 \sim \ve^2 S_3$. Combining this with the following expression derived from Eqs.~\eqrefnoEq{chi_def}, \eqrefnoEq{wave_numbers_sl}, and \eqrefnoEq{etaOTiDef},
 \bal
 \etaOTi \left( \dfrac{4}{3} \xc^{\prime 2} - \xs^{\prime 2} \right) = - \chip \rhoOTi \xs^2,
 \eal
the leading-order expansions in $\ve$ of the determinants  $D$ and $D(A_0)$ in \eqref{D_DA0} become
\bsub
\bal
D &= \dfrac{\ii}{\xc} \left( - \chip \rhoOTi \dfrac{\xs^2}{\xt^2} S_2 + \dfrac{\bcp}{\bt} S_3 \right) , \\
D(A_0) &= \xs^2 \dfrac{\bcp}{\bt} \left( - 1 + \chip \rhoOTi \dfrac{\bc}{\bcp} \right) S_1 \nn \\
& \quad + \dfrac{\xc^2}{3} \dfrac{\xs^2}{\xt^2} \left( \dfrac{\xc^{\prime 2}}{\xc^2} - \chip \rhoOTi \right) S_2 \nn \\
& \quad + \dfrac{\xc^2}{3} \dfrac{\bcp}{\bt} \left( 1 - \dfrac{\bc}{\bcp} \dfrac{\xc^{\prime 2}}{\xc^2} \right) S_3.
\eal
\esub
From this and \eqref{f0_general_app}, we obtain the monopole scattering coefficient $\fOsl$  for a thermoelastic particle suspended in a thermoviscous fluid,
\begin{widetext}
\bal
\eqlab{f0_prelim1}
\fOsl = \frac{3\ii}{\xc^3}\:A_0  = \dfrac{1- \dfrac{1}{\chip \rhoOTi} \dfrac{\xc^{\prime 2}}{\xc^2} - \dfrac{1}{\chip \rhoOTi} \dfrac{\bcp}{\bt} \dfrac{\xt^2}{\xs^2} \left[\dfrac{3 \xs^2}{\xc^2} \left( -1 + \chip \rhoOTi \dfrac{\bc}{\bcp} \right) \dfrac{S_1}{S_2} + \left( 1- \dfrac{\bc}{\bcp} \dfrac{\xc^{\prime 2}}{\xc^2} \right) \dfrac{S_3}{S_2} \right] }{1 - \dfrac{1}{\chip \rhoOTi} \dfrac{\bcp}{\bt} \dfrac{\xt^2}{\xs^2} \dfrac{S_3}{S_2}} .
\eal
\end{widetext}
From \eqref{functions_S_reduced} we obtain the leading-order expansions in $\ve$ for the ratios $S_1/S_2$ and $S_3/S_2$,
\bal
\dfrac{S_1}{S_2} = \left( 1 - \dfrac{1}{\tilde{k}_\mathrm{th}} \dfrac{\bt}{\btp} \right) H(\xt,\xtp) , \quad
\dfrac{S_3}{S_2} = 4 \dfrac{\etaOTi}{\tilde{k}_\mathrm{th}} \dfrac{\bt}{\btp} H(\xt,\xtp),
\eal
with the function $H(\xt,\xtp)$ defined in \eqref{monopole_H}. Inserting this into \eqref{f0_prelim1} and using \eqsref{b_ratios}{etaOTiDef}, and the expression \eqrefnoEq{chi_def} for $\chi'$, we arrive at the final analytical form for  $\fOsl$ given in \eqref{f0_fluid-solid}.

\subsection{The dipole scattering coefficient $\bm{\fI}$}
\seclab{dipole_calculations}
In the long-wavelength limit, for each order $n$, the terms containing $B_n$ and $B_n^\prime$, and thus the variables $\xt$ and $\xtp$, in the system of boundary equations \eqrefnoEq{explicit_bc_eqs_general} are of negligible order relative to the terms containing $A_n$, $A'_n$, $C_n$, $C'_n$, and the inhomogeneous terms. Formally, this is seen by writing up and inverting the entire 6-by-6 matrix equation for the six coefficients for a given $n$. A quicker way to see this, is to write \eqsref{bc_eqs_c}{bc_eqs_d} as
 \bal
 \eqlab{BnBnp}
 & \left( \begin{array}{cc} h_n(\xt) & -j_n(\xtp) \\ \xt h'_n(\xt) & -\xtp j'_n(\xtp) \end{array} \right)
 \left( \begin{array}{c} B_n \\ B'_n \end{array} \right)
 \nn \\
 &\qquad  \sim
 \ve^2 \left( \begin{array}{c} A'_n j_n(\xtp) - A_n h_n(\xc) -j_n(\xc) \\
  A'_n j_n(\xtp) - A_n h_n(\xc) -j_n(\xc) \end{array} \right),
 \eal
where we have used $\frac{b_c}{b_t} , \frac{b'_c}{b_t} \sim \ve^2$ and $\frac{b'_t}{b_t} , \frac{\kth'}{\kth} \sim 1$. Inserting the expressions for $B_n$ and $B'_n$ obtained by inversion of this equation into Eqs.~\eqrefnoEq{bc_eqs_a}, \eqrefnoEq{bc_eqs_b}, \eqrefnoEq{bc_eqs_e}, and \eqrefnoEq{bc_eqs_f}, we see that due to the factor $\ve^2$ each term related to $B_n$ or $B'_n$ are negligible in all four equations. In treating \eqref{bc_eqs_e} it might be useful to use the Bessel's equation~\eqrefnoEq{Bessels_Eq}. Consequently, returning to the dipole problem with $n=1$, terms with $B_1,B_1^{\prime}$ are omitted and the system of equations reduces to four equations with four unknowns, namely \eqref{bc_eqs_a}, \eqref{bc_eqs_b}, \eqref{bc_eqs_e}, and \eqref{bc_eqs_f} without the terms of $B_1,B_1^{\prime}$. For $n=1$ we thus obtain the simplified system of equations
 \bsub
 \eqlab{n1s}
 \bal
 & \xc j_1^{\prime}(\xc) + A_1 \xc h_1^{\prime}(\xc) - 2 C_1 h_1(\xs) \nn \\
 &\quad\quad = A_1^{\prime} \xcp j_1^{\prime}(\xcp) - 2 C_1^{\prime} j_1(\xsp) , \eqlab{n1s_a}
 \\
 & j_1(\xc) + A_1 h_1(\xc) - C_1 \left[ \xs h_1^{\prime}(\xs) + h_1(\xs) \right] \nn \\
 &\quad\quad = A_1^{\prime} j_1(\xcp) - C_1^{\prime} \left[ \xsp j_1^{\prime}(\xsp) + j_1(\xsp)\right] , \eqlab{n1s_b}
 \\
 & \eta_0 \left[ \xc j_2(\xc) + A_1 \xc h_2(\xc) + \frac12 C_1 \xs^2 h_1^{\prime \prime}(\xs) \right] \nn \\
 &\quad\quad = \eta_0^{\prime} \left[A_1^{\prime} \xcp j_2(\xcp) + \frac12 C_1^{\prime} \xs^{\prime 2} j_1^{\prime\prime}(\xsp) \right] , \eqlab{n1s_c}
 \\
 & \eta_0 \left[ \xs^2 j_1(\xc) - 4 \xc j_2(\xc) \right]  - 4 C_1 \eta_0 \xs h_2(\xs) \nn \\
 &\quad\quad\quad + A_1 \eta_0 \left[ \xs^2 h_1(\xc) - 4 \xc h_2(\xc) \right] \nn \\
 &\quad\quad = A_1^{\prime} \eta_0^{\prime} \left[ \xs^{\prime 2} j_1(\xcp) - 4 \xcp j_2(\xcp) \right] \nn \\
 &\quad\quad\quad - 4 C_1^{\prime} \eta_0^{\prime} \xsp j_2(\xsp) , \eqlab{n1s_d}
 \eal
\esub
where we have rewritten the last two equations using the recurrence relations obtained from \eqref{bessel_rec}
\bsub
\bal
x g_1^{\prime}(x) - g_1(x) &= - x g_2(x) , \eqlab{x1}\\
g_1^{\prime\prime}(x) &= - g_1(x) + \frac{2}{x} g_2(x) ,\eqlab{x2}
\eal
\esub
valid for any spherical Bessel or Hankel function $g$.

Simplifying the system of equations we multiply \eqref{n1s_a} by $(-1)$ and add to it \eqref{n1s_b}, then use the recurrence relation \eqrefnoEq{x1}. \eqref{n1s_b} is multiplied by 2 and \eqref{n1s_a} is added while using the recurrence relation
$x g_1^{\prime}(x) + 2 g_1(x) = x g_0(x)$. We leave \eqref{n1s_c} as it is. To \eqref{n1s_d} we add 4 times \eqref{n1s_c} and use the recurrence relation \eqrefnoEq{x2}. With some rearrangements, these manipulations give
 \bsub
 \eqlab{n1ss}
 \bal
 & A_1 \xc h_2(\xc) + C_1 \xs h_2(\xs) \nn \\
 &\quad\quad - A_1^{\prime} \xcp j_2(\xcp) - C_1^{\prime} \xsp j_2(\xsp) = - \xc j_2(\xc) ,
 \\
 & A_1 \xc h_0(\xc) - 2 C_1 \xs h_0(\xs) \nn \\
 &\quad\quad - A_1^{\prime} \xcp j_0(\xcp) + 2 C_1^{\prime} \xsp j_0(\xsp) = - \xc j_0(\xc) ,
 \\
 & A_1 \xc h_2(\xc) + \frac12 C_1 \xs^2 h_1^{\prime\prime}(\xs) \nn \\
 &\quad\quad - \etaOTi \left[ A_1^{\prime} \xcp j_2(\xcp) + \frac12 C_1^{\prime} \xs^{\prime 2} j_1^{\prime\prime}(\xsp)\right] = - \xc j_2(\xc) ,
 \\
 & A_1 h_1(\xc) - 2 C_1 h_1(\xs) \nn \\
 &\quad\quad - \rhoOTi \left[ A_1^{\prime} j_1(\xcp) - 2 C_1^{\prime} j_1(\xsp) \right] = - j_1(\xc) .
 \eal
 \esub
where $\etaOTi \xs^{\prime 2} = \rhoOTi \xs^2$ was used to simplify the last equation. The equations may be further simplified using the relevant scalings in the long-wavelength limit for the fluid droplet and the solid particle, respectively.

\subsubsection{$\fI$ for a suspended thermoviscous droplet}
In the long-wavelength limit for the fluid droplet case the scalings of \eqref{fluid-fluid_longwave} apply. Using the approximate expressions for the spherical Bessel and Hankel functions \eqref{bessel_approx} applicable for small arguments and examining the resulting system of equations \eqrefnoEq{n1ss} one finds that some terms may be omitted to first order in $\ve$. The simplified system of equations \eqrefnoEq{n1ss} for the fluid droplet case takes the form
\bsub
\eqlab{n1sss}
\bal
- \dfrac{3\ii}{\xc^2} A_1 + C_1 \xs h_2(\xs) - C_1^{\prime} \xsp j_2(\xsp) = 0 , \eqlab{n1sss_a}
\eal
\bal
- 2 C_1 \xs h_0(\xs) - A_1^{\prime} \xcp + 2 C_1^{\prime} \xsp j_0(\xsp) = - \xc , \eqlab{n1sss_b}
\eal
\bal
- \dfrac{3\ii}{\xc^2} A_1 + \dfrac{1}{2} C_1 \xs^2 h_1^{\prime\prime}(\xs) - \dfrac{1}{2} C_1^{\prime} \etaOTi \xs^{\prime 2} j_1^{\prime\prime}(\xsp) = 0 , \eqlab{n1sss_c}
\eal
\bal
\dfrac{3\ii}{\xc^2} A_1 + 6 C_1 h_1(\xs) + A_1^{\prime} \rhoOTi\xcp - 6 C_1^{\prime} \rhoOTi j_1(\xsp) = \xc , \eqlab{n1sss_d}
\eal
\esub
Subtracting \eqref{n1sss_c} from \eqref{n1sss_a} and using \eqref{x2}, we can express $C_1^{\prime}$ by $C_1$,
\bsub
\bal
C_1^{\prime} &= \dfrac{\xs^2 h_1(\xs)}{\etaOTi \xsp Q(\xsp)} C_1 , \eqlab{C1p} \\
Q(\xsp) &= \xsp j_1(\xsp) - 2 (1-\fracsmall{1}{\etaOTi}) j_2(\xsp). \eqlab{Qfunc}
\eal
\esub
Then, using this relation to eliminate $C_1^{\prime}$ in \eqref{n1sss_a}, we arrive at the first of the two equations in \eqref{n1_2eq}. The second equation~\eqrefnoEq{n1_2eq_b} is obtained by adding \eqref{n1sss_b} and \eqref{n1sss_d} in order to eliminate $A_1^{\prime}$, then making use of the recurrence relation $3 g_1(x) - x g_0(x) = x g_2(x)$. The resulting two equations for $A_1$ and $C_1$ are
 \bsub
 \eqlab{n1_2eq}
 \bal
 \eqlab{n1_2eq_a}
 \dfrac{3\ii}{\xc^2} A_1 &- C_1 \left[\xs h_2(\xs) - \dfrac{\xs^2 h_1(\xs) j_2(\xsp)}{\etaOTi Q(\xsp) } \right] = 0,
 \eal
 \bal
 \dfrac{3\ii}{\xc^2} A_1 &+ 2 C_1 \rhoOTi \!\! \left[ \frac{3}{\rhoOTi} h_1(\xs) -  \xs h_0(\xs) - \dfrac{\xs^2 h_1(\xs) j_2(\xsp)}{\etaOTi Q(\xsp)} \right]
 \nn \\
 \eqlab{n1_2eq_b}
 &= (1-\rhoOTi) \xc .
 \eal
 \esub
\begin{widetext}
From this, and using again the relation $3 g_1(x) - x g_0(x) = x g_2(x)$, we obtain the dipole expansion coefficient $A_1$,
 \bal
 A_1 &= \dfrac{
 \frac{\ii}{3} \xc^3 (\rhoOTi - 1) \big[ h_2(\xs) \etaOTi Q(\xsp) - \xs h_1(\xs) j_2(\xsp) \big]
 }{
 \big[3 h_2(\xs)-2(\rhoOTi-1) h_0(\xs) \big] \etaOTi Q(\xsp) - (2\rhoOTi+1) \xs h_1(\xs) j_2(\xsp)
 } . \eqlab{dipole_A1}
 \eal
 \end{widetext}
This result, but with a small error in the numerator, was first obtained by Epstein and Carhart \cite{Epstein1953}. We reduce the fraction by $\etaOTi Q(\xsp)h_0(\xs)$ and use the explicit expressions for the Bessel and Hankel functions in \eqref{bessel_explicit} to introduce the functions $G(\xs)$ and $F(\xs,\xsp)$ given explicitly in \eqsref{G_def}{F_def}, respectively,
 \bsub
 \bal
 \eqlab{G_def_appendix}
 G(\xs) &= 1 + \frac{h_2(\xs)}{h_0(\xs)} , \\
 F(\xs,\xsp) &= \frac{\xs h_1(\xs) j_2(\xsp)}{ \etaOTi h_0(\xs) Q(\xsp)} .
 \eal
 \esub
Then, using that $\fI = -6\ii\:\xc^{-3} A_1$, we arrive at the final expression~\eqrefnoEq{f1_fl_final} for the dipole scattering coefficient $f_1^\mathrm{fl}$.

\subsubsection{$\fI$ for a suspended thermoelastic particle}
In the long-wavelength limit for the solid particle the scalings of \eqref{particle-fluid_longwave} apply. Using the approximate expressions for the spherical Bessel and Hankel functions \eqref{bessel_approx} applicable for small arguments and examining the resulting system of equations \eqrefnoEq{n1ss} one finds that some terms may be omitted to first order in $\ve$. The simplified system of equations \eqrefnoEq{n1ss} in the solid particle case takes the form
 \bsub
 \eqlab{n1sss_sl}
 \bal
 &- \dfrac{3\ii}{\xc^2} A_1 + C_1 \xs h_2(\xs) = 0 , \eqlab{n1sss_a_sl}
 \\
 &- 2 C_1 \xs h_0(\xs) - A_1^{\prime} \xcp + 2 C_1^{\prime} \xsp = - \xc , \eqlab{n1sss_b_sl}
 \\
 &- \dfrac{3\ii}{\xc^2} A_1 + \dfrac{1}{2} C_1 \xs^2 h_1^{\prime\prime}(\xs) - \dfrac{1}{15} A_1^\prime \etaOTi \xc^{\prime 3}
 + \dfrac{1}{10} C_1^{\prime} \etaOTi \xs^{\prime 3} = 0 , \eqlab{n1sss_c_sl}
 \\
 &- \dfrac{3\ii}{\xc^2} A_1 - 6 C_1 h_1(\xs) - A_1^{\prime} \rhoOTi \xcp + 2 C_1^{\prime} \rhoOTi \xsp = - \xc , \eqlab{n1sss_d_sl}
 \eal
 \esub
Multiplying \eqref{n1sss_b_sl} by $(-\tilde{\rho_0})$ and adding it to \eqref{n1sss_d_sl}, then substituting $C_1$ using \eqref{n1sss_a_sl}, and finally using the recurrence relation $3 g_1(x) - x g_0(x) = x g_2(x)$, leads to the expansion coefficient $A_1$,
\bal
A_1 &= \dfrac{\frac{\ii}{3} \xc^3 (\rhoOTi - 1) h_2(\xs)}{3 h_2(\xs)-2(\rhoOTi-1) h_0(\xs)} .
\eal
Again, using that $\fI = -6\ii\xc^{-3}\:A_1$ and introducing $G(\xs)$ as defined in \eqref{G_def_appendix}, we obtain after some rearrangement the final result for $f_1^\mathrm{sl}$ given in \eqref{f1_fluid-solid}.\\[5mm]

\section{Special functions}
\seclab{special_funcs}
The Legendre differential equation solved by Legendre polynomials $P_n(\cos\theta)$ of order $n$ is \cite{Arfken2005}
\bal
\eqlab{legendre_eq}
\dfrac{1}{\sin\theta} \dfrac{\mathrm{d} \: }{\mathrm{d}\theta} \left( \sin\theta \dfrac{\mathrm{d} \: }{\mathrm{d}\theta} P_n(\cos\theta) \right) + n(n+1) P_n(\cos\theta) = 0 .
\eal

The Bessel differential equation solved by spherical Bessel or Hankel functions $g_n(x)$ of order $n$ is \cite{Arfken2005}
 \beq{Bessels_Eq}
 x^2\big[g''_n(x) + g_n(x)\big] = n (n+1)g_n(x) -2xg'_n(x) ,
 \eeq
with a prime indicating differentiation with respect to the argument. Useful recurrence relations for $g_n(x)$ are
 \bsub
 \eqlab{bessel_rec}
 \bal
 \eqlab{bessel_rec_a}
 \dfrac{\mathrm{d}}{\mathrm{d}x} \left[ x^{-n} g_n(x) \right] &= - x^{-n} g_{n+1}(x) ,
 \\
 \eqlab{bessel_rec_b}
 \dfrac{\mathrm{d}}{\mathrm{d}x} \left[ x^{n+1} g_n(x) \right] &= x^{n+1} g_{n-1}(x) .
 \eal
 \esub
The lowest-order spherical Bessel functions $j_n(x)$ and Hankel functions of the first kind $h_n(x)$ are \cite{Arfken2005}
 \bsub
 \eqlab{bessel_explicit}
 \bal
 j_0(x) &= \dfrac{\sin x}{x} , \quad
 j_1(x) = \dfrac{1}{x} \left(\dfrac{\sin x}{x} - \cos x \right) , \\
 j_2(x) &= \dfrac{1}{x}\left[\left(\dfrac{3}{x^2}-1\right) \sin x - \dfrac{3}{x} \cos x\right] , \\
 h_0(x) &= - \ii \dfrac{\mathrm{e}^{\ii x}}{x} , \quad
 h_1(x) = - \dfrac{\mathrm{e}^{\ii x}}{x} \left( 1 + \dfrac{\ii}{x}\right) , \\
 h_2(x) &= \ii \dfrac{\mathrm{e}^{\ii x}}{x} \left( 1 + \dfrac{3\ii}{x} - \dfrac{3}{x^2}\right) .
 \eal
 \esub
For small arguments, $x \ll 1$, to first order
\bsub
\eqlab{bessel_approx}
\begin{alignat}{3}
j_0(x) &\simeq 1 , \quad &
j_0^{\prime}(x) &\simeq - \dfrac{x}{3} , \quad &
j_0^{\prime\prime}(x) &\simeq - \dfrac{1}{3} , \\
h_0(x) &\simeq 1 - \dfrac{\ii}{x} , \quad &
h_0^{\prime}(x) &\simeq \dfrac{\ii}{x^2} , \quad &
h_0^{\prime\prime}(x) &\simeq - \dfrac{2\ii}{x^3} , \\
j_1(x) &\simeq \dfrac{x}{3} , \quad &
j_2(x) &\simeq \dfrac{x^2}{15} , & \\
h_1(x) &\simeq - \dfrac{\ii}{x^2} , \quad &
h_2(x) &\simeq - \dfrac{3\ii}{x^3} . &
\end{alignat}
\esub
%

%%%%%%%%%%%%%%%%%%%%%%%%%%%%%%%%%%%%%%%%%%%%%%%%%%%%%%%%%%%%%%%%%%%
%
% Bibliography
%
%%%%%%%%%%%%%%%%%%%%%%%%%%%%%%%%%%%%%%%%%%%%%%%%%%%%%%%%%%%%%%%%%%%

%\bibliography{acoustofluidics}
%\bibliography{library}


\end{document}